\documentclass{article}

\usepackage[utf8]{inputenc}
\usepackage[T1]{fontenc}
\usepackage{amsmath}
\usepackage{amsfonts}
\usepackage{graphicx}
\usepackage{caption}
\usepackage{booktabs}
\usepackage{nicefrac}
\usepackage{microtype}
\usepackage{lipsum}
\usepackage{natbib}
\usepackage{doi}
\usepackage{hyperref}
\usepackage[a4paper, margin=1in]{geometry}

\title{
  {\small \textbf{PREPRINT:} This is a preprint of the paper accepted by the International Conference on Evaluation and Assessment in Software Engineering (EASE25) – AI Models and Data Evaluation Track.} \\[0.5em]
  Benchmarking LLM for Code Smells Detection: OpenAI GPT-4.0 vs DeepSeek-V3
}

\author{
  \href{https://orcid.org/0000-0001-8291-2211}{Ahmed R. Sadik}\thanks{Honda Research Institute Europe, Offenbach am Main, Germany. \texttt{ahmed.sadik@honda-ri.de}} \and
  \href{https://orcid.org/0009-0002-6173-0667}{Siddhata Govind}\thanks{Honda Research Institute Europe, Offenbach am Main, Germany. \texttt{siddhata.govind@honda-ri.de}}
}

\hypersetup{
  pdftitle={Code Smell Detection with LLMs: A Benchmark of GPT-4.0 vs DeepSeek-V3},
  pdfauthor={Ahmed R. Sadik, Siddhata Govind},
  pdfkeywords={Code Smell Detection, Large Language Models, GPT-4.0, DeepSeek-V3},
}

\begin{document}

\maketitle

\setcounter{footnote}{0}

\begin{figure}[!htb]
\centering
\includegraphics[width=0.85\linewidth]{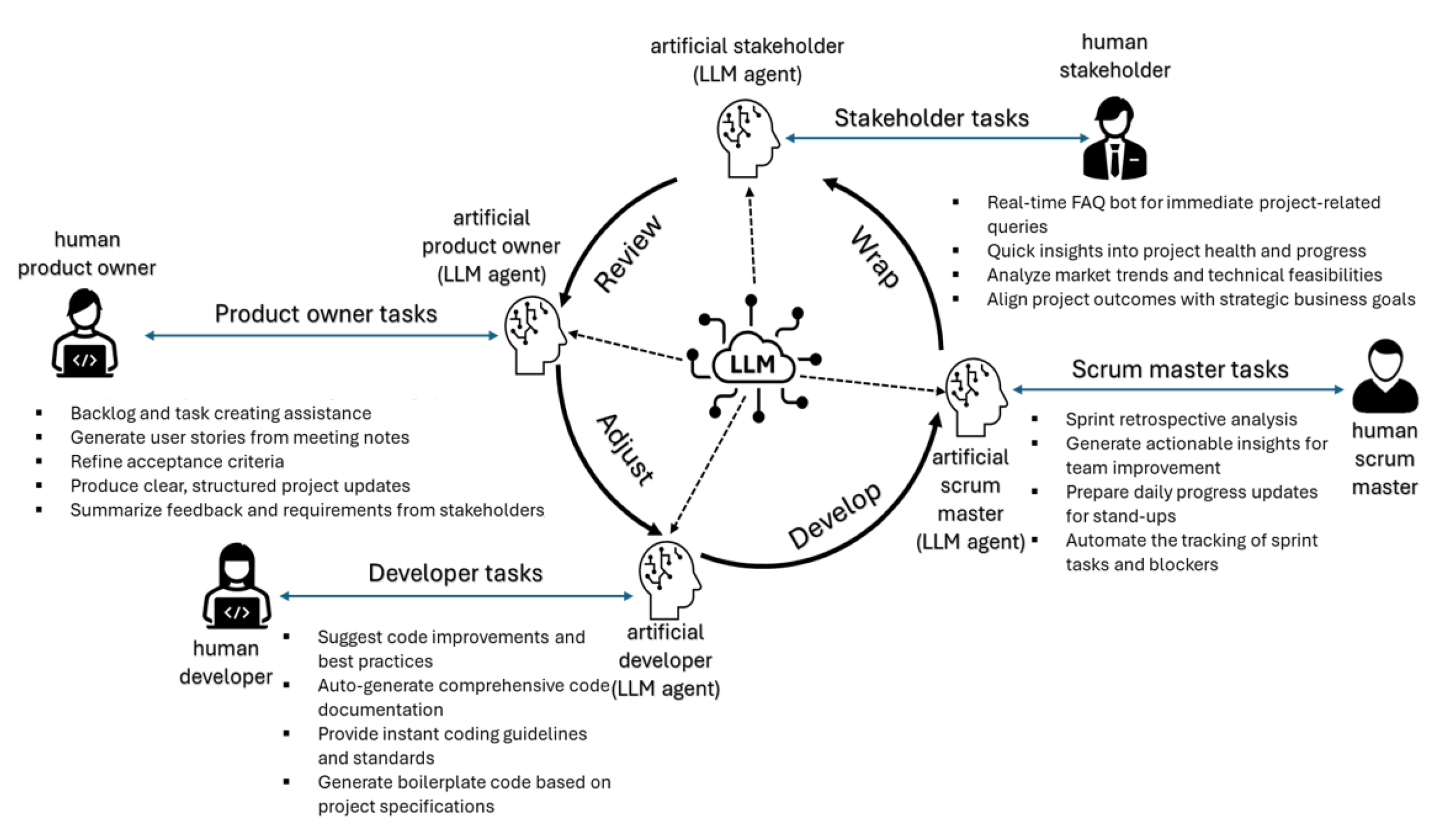}
\caption{LLM-based collaboration in software engineering.}
\label{fig:LLM_Agents}
\end{figure}

\begin{abstract}
Determining the most effective Large Language Model (LLM) for code smell detection presents a complex challenge. This study introduces a structured methodology and evaluation matrix to tackle this issue, leveraging a curated dataset of code samples consistently annotated with known smells. The dataset spans four prominent programming languages—Java, Python, JavaScript, and C++—allowing for cross-language comparison. We benchmark two state-of-the-art LLMs, OpenAI GPT-4.0 and DeepSeek-V3, using precision, recall, and F1-score as evaluation metrics. Our analysis covers three levels of detail: overall performance, category-level performance, and individual code smell type performance. Additionally, we explore cost-effectiveness by comparing the token-based detection approach of GPT-4.0 with the pattern-matching techniques employed by DeepSeek-V3. The study also includes a cost analysis relative to traditional static analysis tools such as SonarQube. The findings offer valuable guidance for practitioners in selecting an efficient, cost-effective solution for automated code smell detection.
\end{abstract}

\begin{center}
\end{center}

\section{Introduction}

The integration of Large Language Models (LLMs) into software engineering is rapidly transforming traditional development workflows by fostering novel forms of collaboration between human developers and artificial intelligence systems \cite{sadik2023coding,sadik2023analysis}. One compelling vision of this transformation is illustrated in Figure~\ref{fig:LLM_Agents}, which shows how LLM-based agents can be embedded across the entire software development lifecycle. These agents augment human roles such as product owners, developers, scrum masters, and stakeholders by providing real-time assistance in task management, code improvement, sprint planning, and requirements clarification \cite{he2024llm, waseem2023chatgpt}.

Each artificial agent, powered by an LLM, mirrors its human counterpart by handling tasks such as auto-generating boilerplate code, analyzing sprint retrospectives, or synthesizing user stories from meeting notes. This human-AI collaboration introduces new efficiencies but also calls for robust, interpretable methods to ensure the quality of outputs—especially in critical areas such as code maintainability. One notable example is the detection and refactoring of code smells: recurring patterns in source code that signal deeper design problems \cite{aranda2024catalog, lucas2024evaluating}. These smells can compromise long-term maintainability, and while traditionally uncovered through manual code reviews or static analysis tools, LLMs now offer a scalable, language-agnostic alternative for automating this task \cite{sadik2023analysis, velasco2024propense}. Motivated by the role of LLMs in these emerging workflows, this paper evaluates the effectiveness of LLM-based agents in detecting code smells across a multilingual dataset. We benchmark the performance of two leading models—GPT-4.0 and DeepSeek-V3—on a shared smelly-code dataset and compare their results against traditional tools like SonarQube \cite{lenarduzzi2020sonarqube, hong2024code}. To offer a detailed assessment, we explore three granularity levels: overall model performance, performance by code smell category, and performance by individual smell type. We further analyze performance across programming languages and evaluate cost-effectiveness.

The remainder of this paper is structured as follows. \textbf{Section~\ref{sec:code_smells}} introduces the taxonomy of code smells, providing detailed definitions and categories that underpin the analysis. \textbf{Section~\ref{sec:dataset}} presents the multilingual dataset developed for this study, highlighting its design, implementation across four programming languages, and associated software metrics. \textbf{Section~\ref{sec:detection_evaluation}} outlines the detection methodology, including prompt design, evaluation metrics, and the experimental setup used to assess LLM performance. \textbf{Section~\ref{sec:language_agnostic_analysis}} provides a language-agnostic analysis, evaluating both models at the overall, category, and type levels. \textbf{Section~\ref{sec:language_analysis}} examines language-specific variations in detection performance across Java, JavaScript, Python, and C++. \textbf{Section~\ref{sec:cost_analysis}} analyzes the cost implications of using GPT-4.0 and DeepSeek-V3 based on pricing models and code complexity. \textbf{Section~\ref{sec:sonarqube_comparison}} compares the capabilities of LLM-based detection with the static analysis tool SonarQube, highlighting key differences in adaptability, explainability, and integration. Finally, \textbf{Section~\ref{sec:discussion_conclusion}} concludes with a discussion of key findings, limitations, and opportunities for future research.

\section{Code Smells}
\label{sec:code_smells}

\begin{figure}[!htbp]
  \centering
  \includegraphics[width=\linewidth]{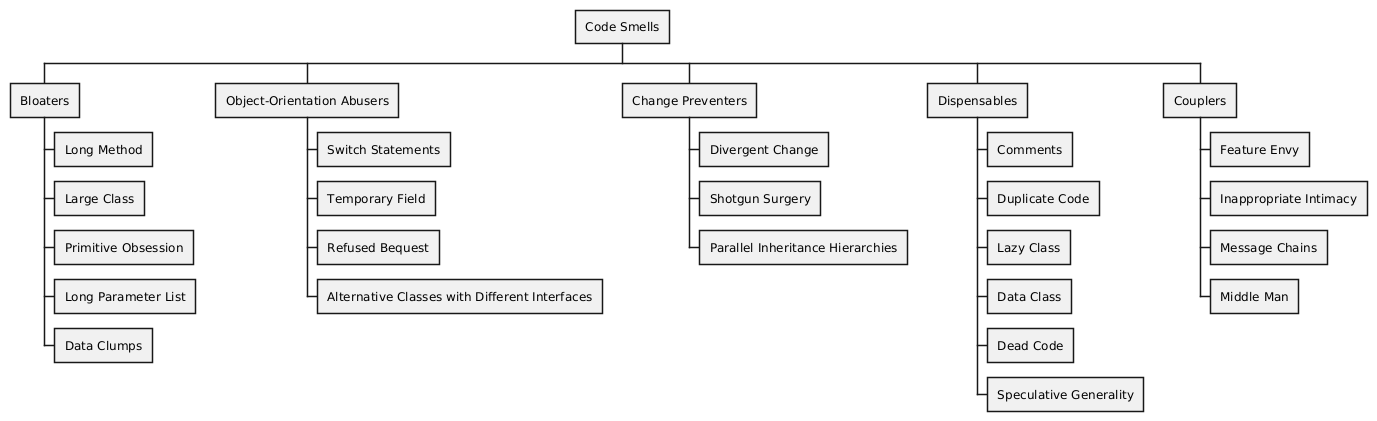}
  \caption{Codesmells Taxonomy.}
  % \Description is not a standard command in the article class
  % \Description{Codesmells Taxonomy.}
\end{figure}

In software engineering, maintaining high code quality is essential for developing robust, maintainable, and scalable systems \cite{sadik2023analysis}. As software projects evolve, they often accumulate design inefficiencies or anomalies, commonly referred to as \textit{code smells} \cite{wu2024ismell}. These code smells serve as indicators of deeper structural issues within the codebase, potentially leading to increased technical debt and maintenance challenges if left unaddressed \cite{lucas2024evaluating}\cite{aranda2024catalog}. Code smells are broadly categorized based on their characteristics and the underlying problems they signify\cite{waseem2023chatgpt}\cite{alves2024detecting}. Understanding these categories is vital for developers and researchers aiming to identify, assess, and remediate such issues effectively\cite{refactoring-guru}. The code smells categories and their associated types include:
\begin{itemize}
    \item \textbf{Bloaters} result from code that has grown excessively, making it difficult to manage and understand. Large classes that contain too many attributes or methods can become difficult to navigate and maintain. Long methods, packed with multiple responsibilities, reduce readability and make debugging harder. Excessive parameter lists in method signatures create confusion and increase the likelihood of errors. Primitive obsession, where primitive data types are used instead of domain-specific objects, leads to convoluted logic and reduces maintainability. Over time, these bloaters degrade code quality and slow down development \cite{marticorena2006extending}.
\end{itemize}

\begin{itemize}
    \item \textbf{Dispensables} refer to unnecessary elements within the code that, if removed, would improve efficiency and clarity. These include redundant comments that may indicate convoluted logic, duplicate code appearing in multiple locations, and classes that exist without contributing meaningful functionality. Additionally, classes containing only fields without behavior add little value, while dead code remains in the system without being executed or used. Speculative generality, where code is written with future use cases in mind but never utilized, also adds unnecessary complexity \cite{gesi2022code}.
\end{itemize}

\begin{itemize}
    \item \textbf{Couplers} involve excessive dependencies between classes, reducing modularity. Feature envy occurs when a method relies more on data from other classes than its own, while inappropriate intimacy results from classes having deep dependencies on the internal workings of other classes. Message chains, where a method calls a sequence of methods across different objects, introduce fragility, and middle-man classes that primarily delegate tasks add unnecessary indirection \cite{tandon2024study}.
\end{itemize}

\begin{itemize}
    \item \textbf{Object-Orientation Abusers} emerge when object-oriented programming principles are either incorrectly applied or insufficiently utilized. These issues manifest in excessive use of switch statements or nested if-else conditions that could be replaced with polymorphism, fields that are occasionally set but remain unused most of the time, and subclasses inheriting methods or data they do not require, leading to unnecessary complexity. Similarly, alternative classes with different interfaces performing similar functionalities introduce redundancy and inconsistency in the codebase \cite{walter2016relationship}.
\end{itemize}

\begin{itemize}
    \item \textbf{Change Preventers} occur when modifying a single aspect of the software necessitates changes in multiple locations, limiting adaptability. This includes cases where a class undergoes different kinds of modifications, leading to poor cohesion, a single change requiring small but widespread adjustments across multiple classes, or parallel inheritance hierarchies that force developers to create corresponding subclasses in multiple locations \cite{haque2018causes}..
\end{itemize} 

In this study, we aim to benchmark the effectiveness of LLMs, specifically GPT-4.0, against specialized tools like DeepSeek in detecting these code smells across various programming languages \cite{sadik2023coding}. By evaluating their performance, we seek to provide insights into the applicability of LLMs in automated code quality assessment and their potential integration into modern software development workflows.

\section{Dataset}
\label{sec:dataset}

\begin{figure}[h]
    \centering
    \includegraphics[width=0.9\linewidth]{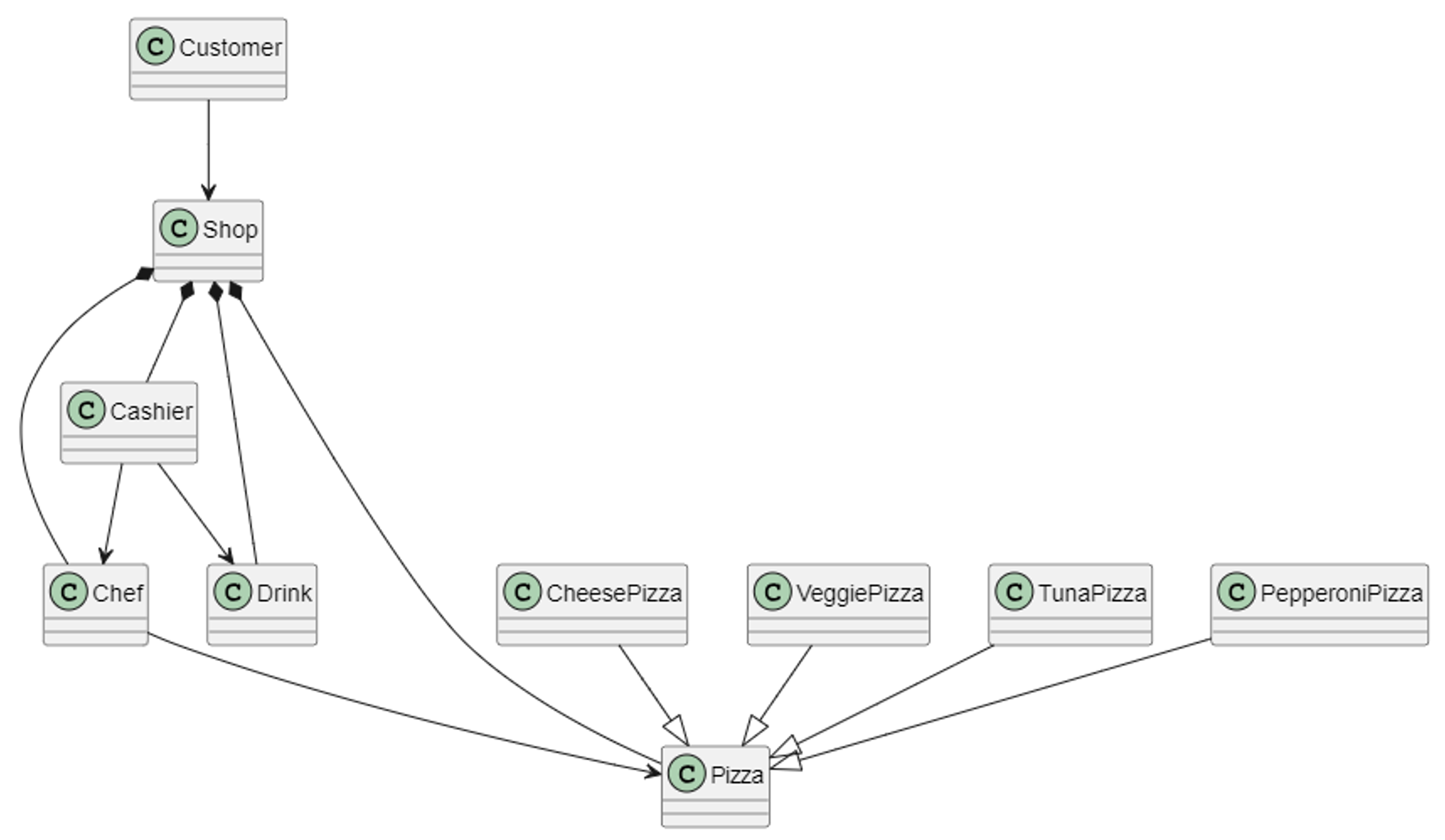}
    \caption{Generic class diagram representing the generic system implemented in Java, JavaScript, Python, and C++.}
    \label{fig:class_diagram}
\end{figure}

\begin{figure}[!htbp]
    \centering
    \includegraphics[width=0.95\linewidth]{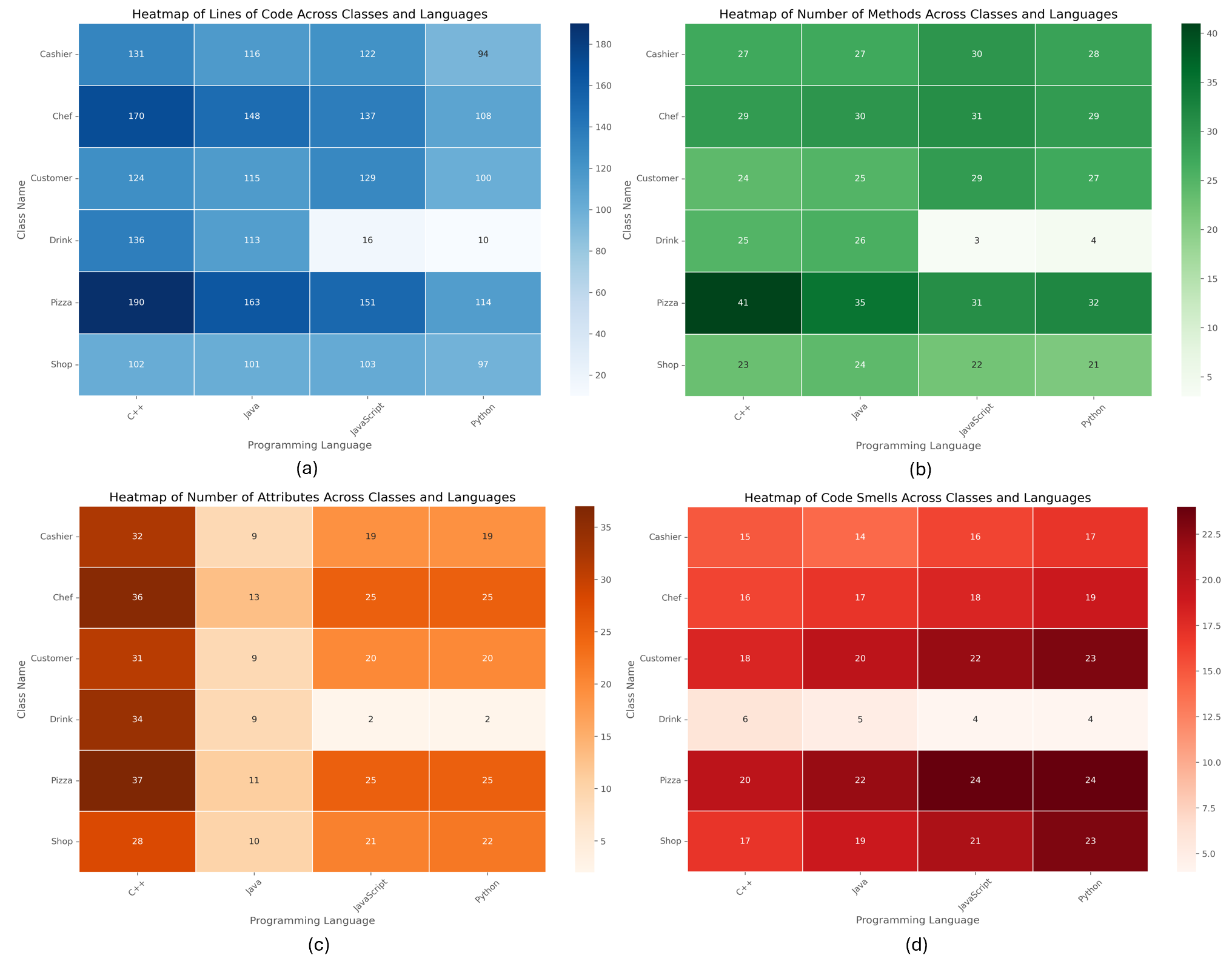}
    \caption{Heatmaps representing various metrics across Java, JavaScript, Python, and C++ implementations. (a) Lines of Code (LOC), (b) Number of Methods, (c) Number of Attributes, (d) Code Smells Distribution.}
    \label{fig:heatmaps}
\end{figure}

During this study, we created a dataset that available on GitHub \cite{sadik2025smellycode} and Zenodo \cite{sadik2025smellycodedataset} contains code with well-known smells to act as the analysis ground truth. The dataset is a generic implementation of a restaurant management system, designed consistently across four programming languages: Java, JavaScript, Python, and C++. Figure~\ref{fig:class_diagram} illustrates the key components and relationships within this system, ensuring a standardized design that allows cross-language comparison. 

At the core of the system is the `Shop` class, which acts as the central entity managing customer interactions, order processing, and coordination with employees. The `Customer` class represents individuals placing orders, which are processed through the `Cashier`. The `Cashier` communicates with the `Chef`, responsible for food preparation, and manages various restaurant items such as `Drink` and `Pizza`. The `Pizza` class serves as a base for multiple pizza variants, including `CheesePizza`, `VeggiePizza`, `TunaPizza`, and `PepperoniPizza`, following an inheritance model to maintain a structured hierarchy. This design reflects core object-oriented principles such as encapsulation and polymorphism while ensuring consistency across all four language implementations. The relationships between these classes model real-world restaurant operations, with direct interactions between customers, employees, and food preparation units. This structured representation provides a foundation for evaluating code quality, identifying design flaws, and analyzing the presence of code smells across different object-oriented programming paradigms.

To build the dataset ground truth, we implement the case study across different programming paradigms, we implemented the restaurant management system in four widely used languages: Java, JavaScript, Python, and C++. This dataset captures key software metrics to compare the structure and design choices across these implementations. The collected data is visualized through heatmaps in Figure~\ref{fig:heatmaps}, which illustrate variations in Lines of Code (LOC), number of methods, attributes, and existing code smells. The heatmap in Figure~\ref{fig:heatmaps}a presents the LOC distribution across the different classes in each language. The `Pizza` class exhibits the highest LOC values across all implementations, suggesting potential instances of the `Large Class` code smell. Similarly, `Chef` and `Cashier` maintain relatively high LOC values, reflecting their central roles in processing orders. The `Drink` class, in contrast, remains consistently small, indicating its limited functionality. Figure~\ref{fig:heatmaps}b displays the number of methods across implementations. The `Pizza` class contains the highest method count, reinforcing the likelihood of `Long Method` or `God Class` smells. The distribution of methods also varies by language, particularly in JavaScript and Python, which generally feature fewer methods per class due to their dynamic nature. The attribute count, visualized in Figure~\ref{fig:heatmaps}c, shows that `Cashier`, `Chef`, and `Pizza` have the most attributes across implementations, indicating a tendency toward high class coupling. The presence of excessive attributes in these classes suggests the potential for `Data Clumps` and `Primitive Obsession` code smells. Conversely, `Drink` and `Customer` exhibit lower attribute counts, implying better modularity. Finally, Figure~\ref{fig:heatmaps}d highlights the distribution of detected code smells across the four implementations. The `Pizza` and `Shop` classes show a higher concentration of code smells, especially in C++ and Java, where static typing often leads to more verbose implementations. Python and JavaScript demonstrate slightly fewer code smells, likely due to their flexibility in handling class structures. However, the `Drink` class remains consistently low in detected code smells, indicating a well-encapsulated design.

This dataset provides a structured and comparative analysis of how different programming languages impact code structure and maintainability. The variations observed in LOC, method counts, attribute counts, and code smells highlight key trade-offs in object-oriented design across static and dynamic typing paradigms. These findings serve as the basis for further evaluation of code quality and the effectiveness of LLMs in detecting and refactoring code smells.

\section{Detection and Evaluation Methods}
\label{sec:detection_evaluation}

To evaluate code smells detection \cite{paiva2017evaluation}, we utilize LLMs by first stripping any manually annotated smells from the case study classes. The cleaned code is then passed to the selected model—either OpenAI GPT-4.0 or DeepSeek-V3—using a standardized prompt that instructs it to detect and categorize code smells across classes, programming languages, and smell categories.

The effectiveness of the detection is assessed by comparing model predictions against the ground truth annotations. We define the following metrics: True Positives (TP) denote correctly identified code smells, False Positives (FP) occur when the model incorrectly flags clean code as smelly, and False Negatives (FN) indicate missed detections.

We evaluate the model using three key performance indicators:

\begin{itemize}
    \item \textbf{Precision} : Measures the accuracy of detected code smells \cite{quba2021software}.
    \begin{math}
    Precision = \frac{TP}{TP + FP}
    \end{math}
    
    \item \textbf{Recall}: Quantifies the model’s ability to identify all actual smells \cite{haque2018causes}.
    \begin{math}
    Recall = \frac{TP}{TP + FN}
    \end{math}
    
    \item \textbf{F1-Score}: Provides a balance between precision and recall \cite{sahin2019conceptual}.
    \begin{math}
    F1 = 2 \times \frac{Precision \times Recall}{Precision + Recall}
    \end{math}
\end{itemize}

\begin{table}[h]
  \caption{Evaluation Metrics for Code Smell Detection}
  \label{tab:evaluation}
  \centering
  \begin{tabular}{lcc}
    \toprule
    Metric & Formula & Description \\
    \midrule
    Precision & $\frac{TP}{TP + FP}$ & Accuracy of detected smells \\
    Recall & $\frac{TP}{TP + FN}$ & Detection coverage \\
    F1-Score & $2 \times \frac{Precision \times Recall}{Precision + Recall}$ & Precision-Recall balance \\
    \bottomrule
  \end{tabular}
\end{table}

Table~\ref{tab:evaluation} summarizes this metric, which provides a structured evaluation framework for comparing LLM performance in detecting code smells across different programming languages.

\section{Language Agnostic Analysis}
\label{sec:language_agnostic_analysis}

\subsection{LLM's Model Level}

Building on the evaluation metrics outlined in the previous section, we analyze the performance of GPT-4.0 and DeepSeek-V3 in detecting code smells across multiple programming languages. This analysis is based on the precision, recall, and F1-score metrics, which provide insight into the effectiveness of each model in identifying and classifying code smells. Table~\ref{tab:model_comparison} summarizes the results, showing that GPT-4.0 achieves significantly higher precision (0.79) than DeepSeek-V3 (0.42). This indicates that GPT-4.0 generates fewer false positives, making it a more reliable model in accurately identifying code smells. However, both models exhibit relatively low recall, with GPT-4.0 achieving 0.41 and DeepSeek-V3 at 0.31, suggesting that a considerable number of actual code smells remain undetected.

\begin{table}[h]
  \caption{Comparison of GPT-4.0 and DeepSeek-V3 in Code Smell Detection}
  \label{tab:model_comparison}
  \centering
  \begin{tabular}{lccc}
    \toprule
    Model & Precision & Recall & F1-score \\
    \midrule
    GPT-4.0 & 0.79 & 0.41 & 0.54 \\
    DeepSeek & 0.42 & 0.31 & 0.35 \\
    \bottomrule
  \end{tabular}
\end{table}

\begin{figure}[!htbp]
    \centering
    \includegraphics[width=\linewidth]{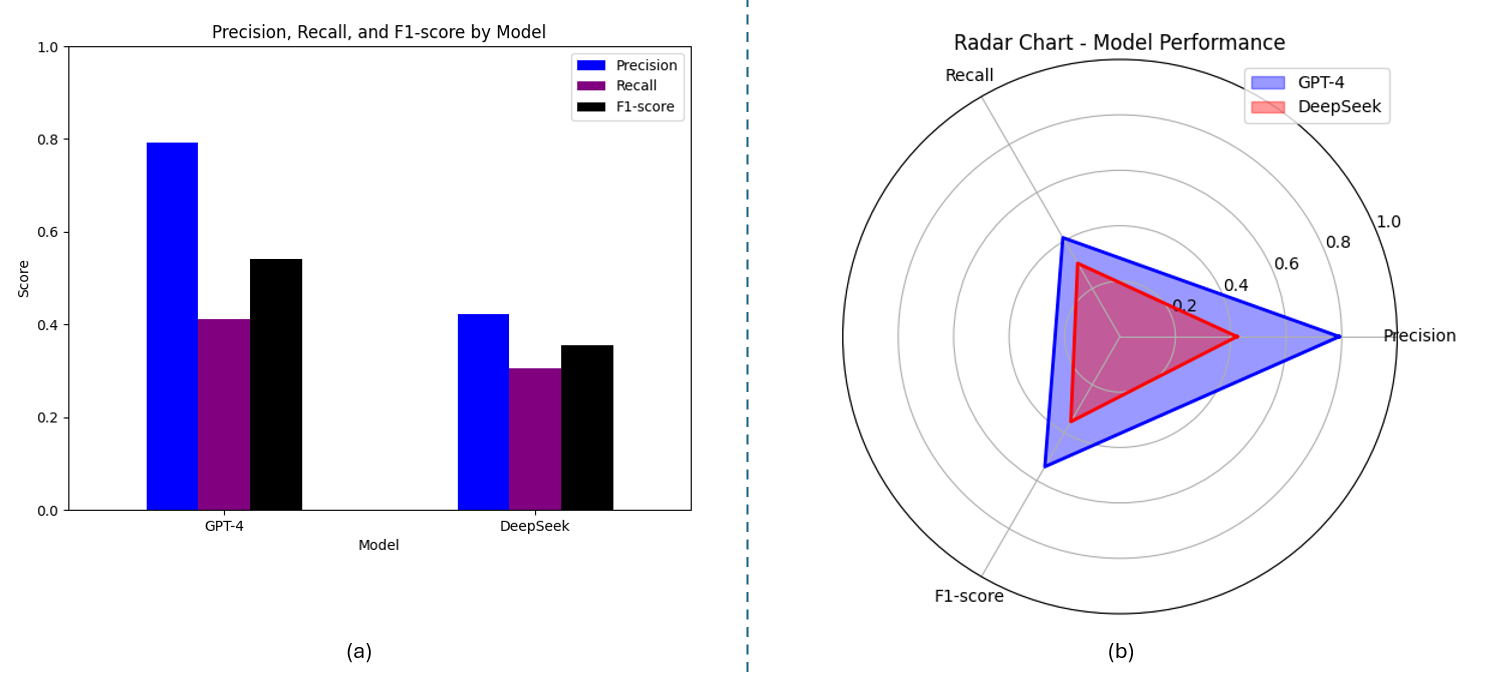}
    \caption{Performance comparison of GPT-4.0 and DeepSeek-V3 in code smell detection.}
    \label{fig:model_level_analysis}
\end{figure}

Figure~\ref{fig:model_level_analysis} provides a visual representation of the models' performance. The left-hand bar chart highlights GPT-4.0's superior precision, while the radar chart on the right illustrates how both models struggle with recall and F1-score. The distribution of true positives, false positives, and false negatives indicates that while GPT-4.0 correctly identifies more actual code smells, DeepSeek-V3 produces a significantly higher number of false positives.

The analysis reveals that GPT-4.0, despite its higher precision, does not capture all instances of code smells, resulting in a lower recall. Conversely, DeepSeek-V3, while identifying a broader range of potential smells, suffers from excessive false positives, reducing its reliability. These findings suggest that while GPT-4.0 is a more precise option, further improvements are needed to enhance recall and overall detection coverage.

\newpage

\subsection{Code Smell's Category level}

\begin{figure}[!htbp]
    \centering
    \includegraphics[width=\linewidth]{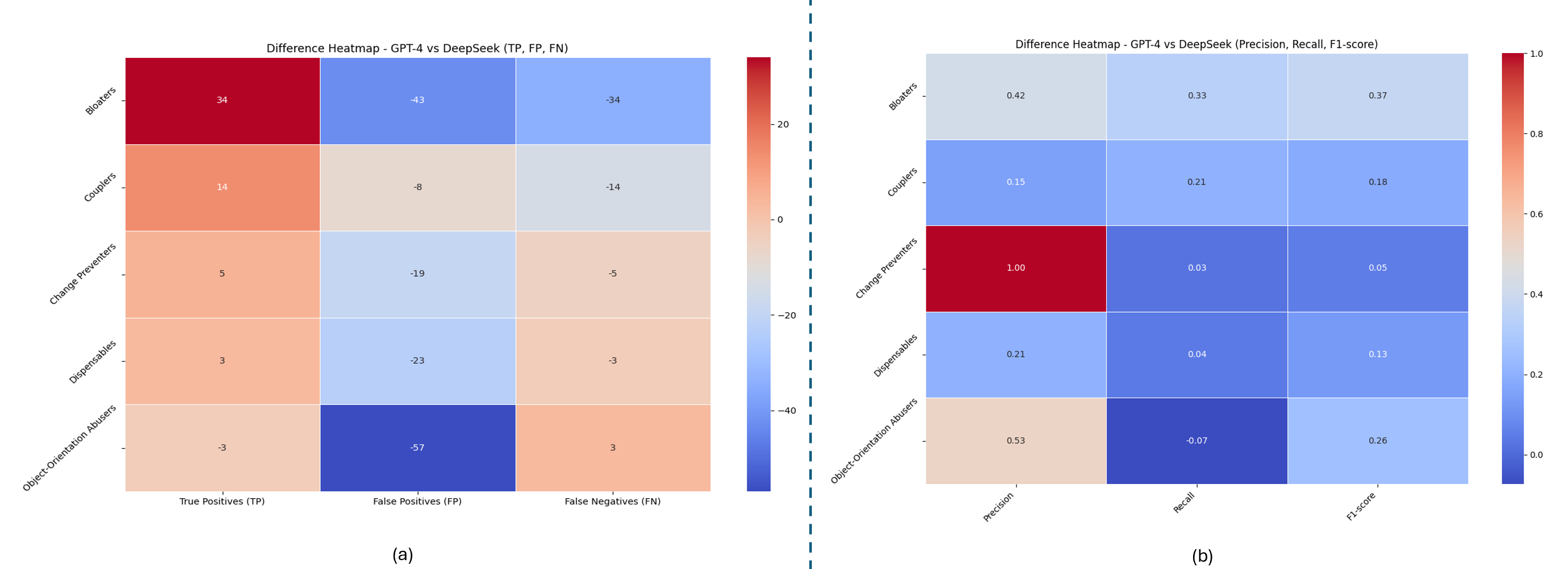}
    \caption{Comparison of GPT-4.0 vs. DeepSeek at the category level: (a) TP, FP, FN difference heatmap, (b) Precision, Recall, and F1-score difference heatmap.}
    \label{fig:category_level_analysis}
\end{figure}

Building on the evaluation metrics outlined in the previous section, we analyze the performance of GPT-4.0 and DeepSeek-V3 at the category level. This analysis focuses on how each model performs in detecting specific types of code smells, measured using precision, recall, and F1-score. Table~\ref{tab:category_comparison} summarizes the results, showing that GPT-4.0 consistently achieves higher precision across all categories, particularly in detecting Change Preventers (1.00) and Object-Orientation Abusers (0.53). This suggests that GPT-4.0 generates fewer false positives and is more reliable in classifying code smells. However, both models exhibit relatively low recall, with GPT-4.0 achieving near-zero recall for Change Preventers and DeepSeek struggling to capture a significant number of actual code smells.

\begin{table}[h]
  \caption{Comparison of GPT-4.0 and DeepSeek-V3 in Code Smell Detection by Category}
  \label{tab:category_comparison}
  \centering
  \begin{tabular}{lccccccc}
    \toprule
    Model & Category & TP & FP & FN & Precision & Recall & F1-score \\
    \midrule
    GPT-4.0 & Bloaters & 62 & 25 & 40 & 0.71 & 0.61 & 0.66 \\
    DeepSeek & Bloaters & 28 & 68 & 74 & 0.29 & 0.27 & 0.28 \\
    GPT-4.0 & Couplers & 58 & 5 & 10 & 0.92 & 0.85 & 0.89 \\
    DeepSeek & Couplers & 44 & 13 & 24 & 0.77 & 0.65 & 0.70 \\
    GPT-4.0 & Change Preventers & 5 & 0 & 190 & 1.00 & 0.03 & 0.05 \\
    DeepSeek & Change Preventers & 2 & 1 & 193 & 0.67 & 0.01 & 0.02 \\
    GPT-4.0 & Dispensables & 18 & 4 & 23 & 0.82 & 0.44 & 0.57 \\
    DeepSeek & Dispensables & 15 & 9 & 26 & 0.63 & 0.37 & 0.47 \\
    GPT-4.0 & Object-Orientation Abusers & 25 & 12 & 28 & 0.68 & 0.47 & 0.56 \\
    DeepSeek & Object-Orientation Abusers & 28 & 30 & 25 & 0.48 & 0.53 & 0.51 \\
    \bottomrule
  \end{tabular}
\end{table}

Figure~\ref{fig:category_level_analysis} provides a visual representation of the models' performance across different categories. The left heatmap illustrates the difference in TP, FP, and FN counts, showing that GPT-4.0 has significantly higher true positives in Bloaters and Couplers, while DeepSeek has a high false positive rate across all categories. The right heatmap highlights GPT-4.0's precision advantage, particularly in Change Preventers and Object-Orientation Abusers, while DeepSeek shows a tendency to over-predict certain smells, leading to inflated false positive rates. The results indicate that while GPT-4.0 is more precise, it still struggles with recall in categories like Change Preventers, where its detection is nearly absent. DeepSeek, while achieving slightly better recall in certain cases, compensates by producing excessive false positives, reducing its reliability. These findings suggest that a hybrid approach combining LLM detection with static analysis tools may be needed to improve recall while maintaining precision. Further research into category-specific tuning could help enhance the models' effectiveness in detecting challenging code smells.

\newpage

\subsection{Code Smell's Type level}

\begin{figure}[!htbp]
    \centering
    \includegraphics[width=\linewidth]{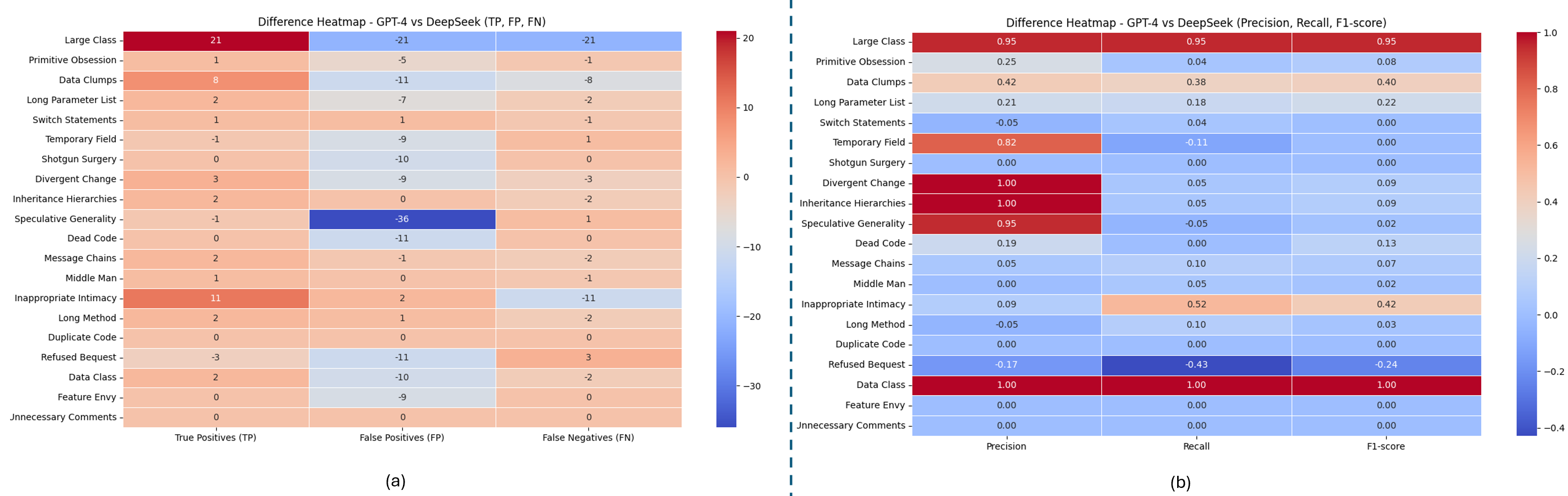}
    \caption{Comparison of GPT-4.0 vs. DeepSeek at the type level: (a) TP, FP, FN difference heatmap, (b) Precision, Recall, and F1-score difference heatmap.}
    \label{fig:type_level_analysis}
\end{figure}

To further analyze the performance of GPT-4.0 and DeepSeek-V3 in detecting code smells, we examine their performance at the type level. This involves comparing true positives (TP), false positives (FP), and false negatives (FN) across various code smell types, as well as evaluating their impact on precision, recall, and F1-score. Table~\ref{tab:type_comparison} presents a detailed comparison between the two models across different code smell types. GPT-4.0 consistently outperforms DeepSeek in detecting certain types, most notably in \textit{Large Class}, where it identifies 21 true positives compared to none for DeepSeek, leading to a significantly higher precision and recall. Similarly, GPT-4.0 exhibits strong performance in detecting \textit{Data Clumps}, \textit{Switch Statements}, and \textit{Inappropriate Intimacy}, demonstrating higher recall and fewer false positives than DeepSeek.While GPT-4.0 tends to have superior precision across most code smells, recall remains a challenge in some cases. For instance, in \textit{Shotgun Surgery} and \textit{Refused Bequest}, both models struggle to detect actual instances, leading to very low recall values. DeepSeek demonstrates a marginal advantage in recall for \textit{Refused Bequest} (0.43) but at the cost of a significantly higher false positive rate (15 FP vs. 4 FP for GPT-4.0). Conversely, GPT-4.0 achieves perfect precision in \textit{Divergent Change}, \textit{Data Class}, and \textit{Message Chains}, while DeepSeek struggles in these areas, generating a large number of false positives.

\begin{table}[h]
  \caption{Comparison of GPT-4.0 and DeepSeek-V3 in Code Smell Detection by Type (Summary)}
  \label{tab:type_comparison}
  \centering
  \begin{tabular}{lccccccc}
    \toprule
    Model & Code Smell & TP & FP & FN & Precision & Recall & F1-score \\
    \midrule
    GPT-4.0 & Large Class & 21 & 1 & 1 & 0.95 & 0.95 & 0.95 \\
    DeepSeek & Large Class & 0 & 22 & 22 & 0.00 & 0.00 & 0.00 \\
    GPT-4.0 & Data Clumps & 8 & 11 & 13 & 0.42 & 0.38 & 0.40 \\
    DeepSeek & Data Clumps & 0 & 22 & 21 & 0.00 & 0.00 & 0.00 \\
    GPT-4.0 & Switch Statements & 21 & 1 & 4 & 0.95 & 0.84 & 0.89 \\
    DeepSeek & Switch Statements & 20 & 0 & 5 & 1.00 & 0.80 & 0.89 \\
    GPT-4.0 & Dead Code & 31 & 6 & 0 & 0.84 & 1.00 & 0.91 \\
    DeepSeek & Dead Code & 31 & 17 & 0 & 0.65 & 1.00 & 0.78 \\
    GPT-4.0 & Message Chains & 21 & 0 & 0 & 1.00 & 1.00 & 1.00 \\
    DeepSeek & Message Chains & 19 & 1 & 2 & 0.95 & 0.90 & 0.93 \\
    GPT-4.0 & Inappropriate Intimacy & 16 & 4 & 5 & 0.80 & 0.76 & 0.78 \\
    DeepSeek & Inappropriate Intimacy & 5 & 2 & 16 & 0.71 & 0.24 & 0.36 \\
    GPT-4.0 & Refused Bequest & 0 & 4 & 7 & 0.00 & 0.00 & 0.00 \\
    DeepSeek & Refused Bequest & 3 & 15 & 4 & 0.17 & 0.43 & 0.24 \\
    GPT-4.0 & Feature Envy & 0 & 1 & 5 & 0.00 & 0.00 & 0.00 \\
    DeepSeek & Feature Envy & 0 & 10 & 5 & 0.00 & 0.00 & 0.00 \\
    \bottomrule
  \end{tabular}
\end{table}

For brevity, Table~\ref{tab:type_comparison} presents a summarized version of the full dataset, which includes 40 rows covering all evaluated code smells. The key findings remain consistent across the dataset: GPT-4.0 tends to generate more accurate classifications with fewer false positives, while DeepSeek produces more false positives and suffers from lower precision. Figure~\ref{fig:type_level_analysis} provides a heatmap visualization of these differences. The left heatmap (a) illustrates the differences in TP, FP, and FN counts, while the right heatmap (b) highlights variations in precision, recall, and F1-score. The results indicate that GPT-4.0 consistently achieves higher precision across most code smell types, particularly in \textit{Large Class} (0.95), \textit{Divergent Change} (1.00), and \textit{Data Class} (1.00). However, recall remains a challenge for GPT-4.0 in specific cases, such as \textit{Refused Bequest} and \textit{Shotgun Surgery}, where both models exhibit poor recall. DeepSeek shows slightly better recall in some cases but at the expense of an excessive number of false positives.

The analysis reveals that while GPT-4.0 is more precise in classifying code smells, its recall varies across different types, particularly for \textit{Refused Bequest} and \textit{Shotgun Surgery}. DeepSeek, on the other hand, demonstrates marginally better recall in some cases but at the cost of excessive false positives, reducing overall reliability. These findings suggest that an optimal strategy could involve a hybrid approach that integrates GPT-4.0's precision with additional techniques—such as static analysis heuristics—to improve recall while maintaining high reliability.

\section{Language based analysis}
\label{sec:language_analysis}

To further assess the performance of GPT-4.0 and DeepSeek-V3, we extend our analysis to the language level, investigating how these models perform across different programming languages. This allows us to identify language-specific variations in code smell detection capabilities. We analyze differences in precision, recall, and F1-score across C++, Java, JavaScript, and Python.Figures~\ref{fig:type_language_precision}, \ref{fig:type_language_recall}, and \ref{fig:type_language_f1} present heatmaps visualizing the performance differences between GPT-4.0 and DeepSeek-V3 for each code smell type across these four languages. The red hues indicate superior performance by GPT-4.0, whereas blue hues highlight cases where DeepSeek performs comparatively better.

\subsection{Precision}

\begin{figure}[!ht]
    \centering
    \includegraphics[width=\linewidth]{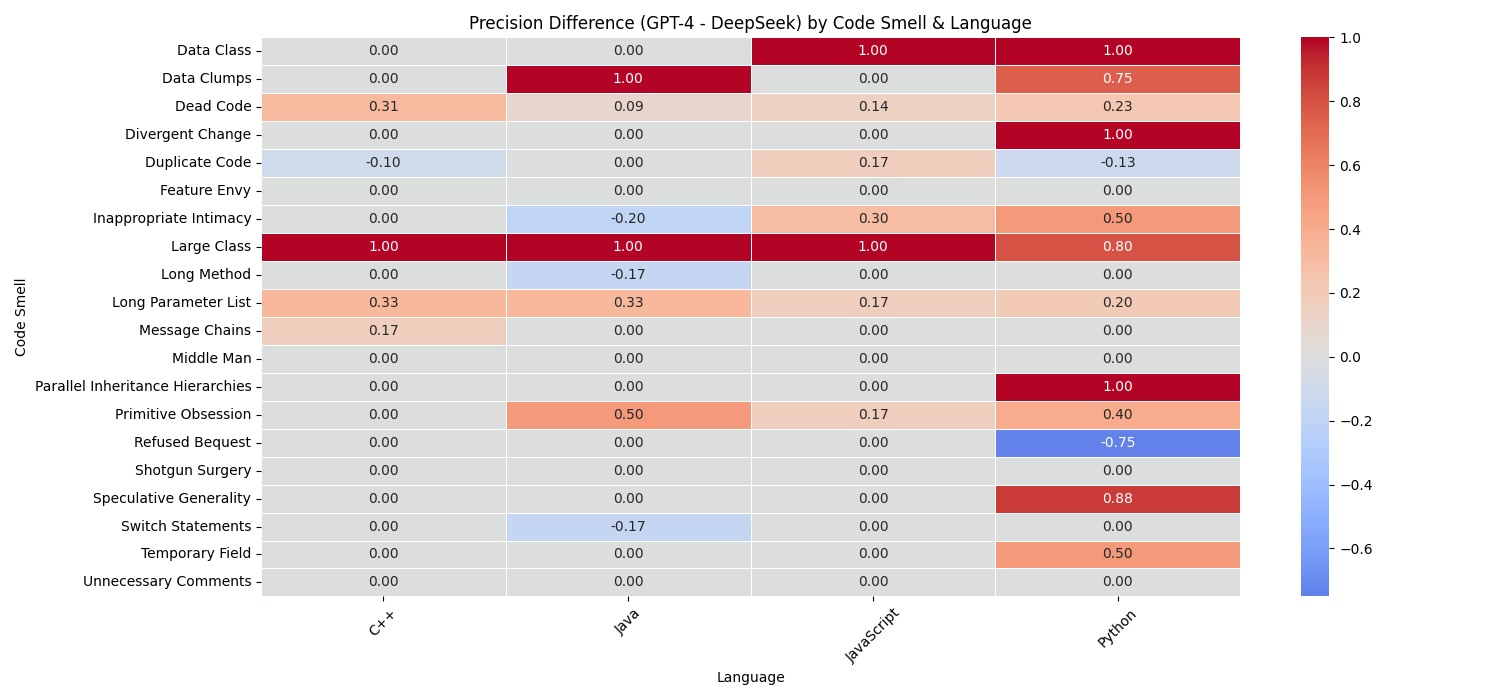}
    \caption{Precision difference (GPT-4.0 - DeepSeek) by code smell and language. Red indicates GPT-4.0's advantage, blue indicates DeepSeek's advantage.}
    \label{fig:type_language_precision}
\end{figure}

Figure~\ref{fig:type_language_precision} illustrates the precision difference, revealing that GPT-4.0 consistently outperforms DeepSeek in most code smells, particularly for \textit{Large Class}, \textit{Divergent Change}, and \textit{Data Class} across all languages. However, some cases show minor weaknesses in GPT-4.0’s precision, such as in \textit{Duplicate Code} in C++ and \textit{Switch Statements} in Java, where DeepSeek produces fewer false positives.

\subsection{Recall}

\begin{figure}[!ht]
    \centering
    \includegraphics[width=\linewidth]{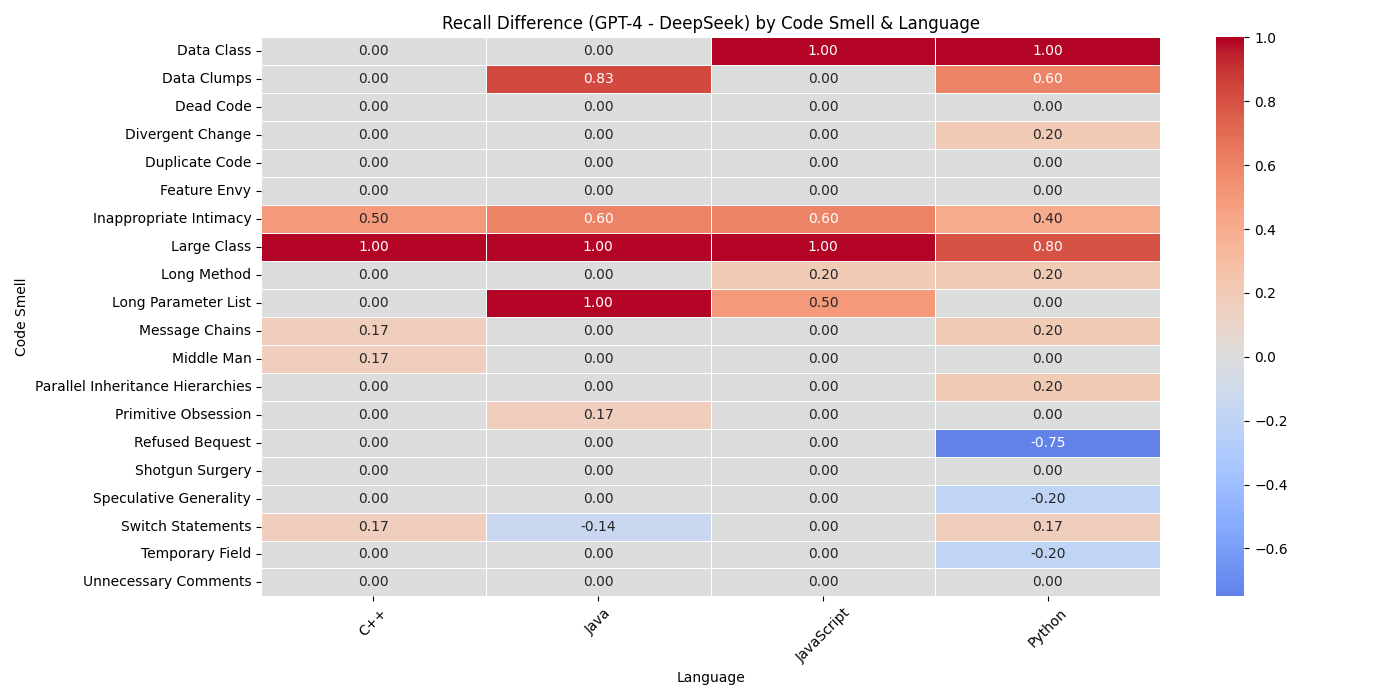}
    \caption{Recall difference (GPT-4.0 - DeepSeek) by code smell and language. Higher recall in GPT-4.0 is shown in red, DeepSeek advantages in blue.}
    \label{fig:type_language_recall}
\end{figure}

Figure~\ref{fig:type_language_recall} presents recall differences across languages. GPT-4.0 exhibits a recall advantage in several cases, particularly in \textit{Large Class}, \textit{Inappropriate Intimacy}, and \textit{Parallel Inheritance Hierarchies}. However, DeepSeek-V3 occasionally surpasses GPT-4.0 in recall, such as in \textit{Refused Bequest} in Python, where it captures more true positives despite a higher false positive rate. This indicates that GPT-4.0’s detection may be more conservative in certain scenarios, missing a number of valid cases.

\subsection{F1-score}

\begin{figure}[!ht]
    \centering
    \includegraphics[width=\linewidth]{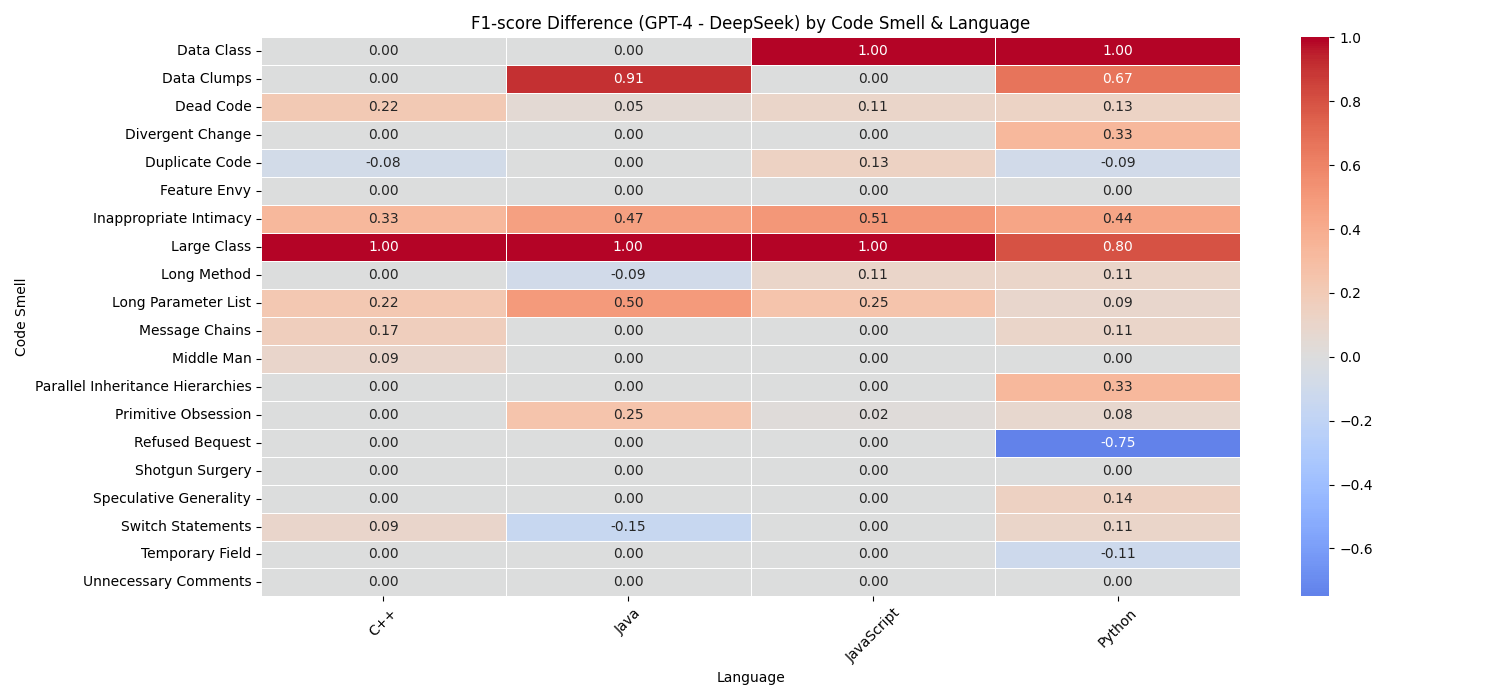}
    \caption{F1-score difference (GPT-4.0 - DeepSeek) by code smell and language. Red areas indicate GPT-4.0’s dominance, blue areas show DeepSeek’s strengths.}
    \label{fig:type_language_f1}
\end{figure}

Finally, Figure~\ref{fig:type_language_f1} visualizes the F1-score difference, which balances precision and recall. The results confirm GPT-4.0's overall superiority in detecting code smells, particularly in \textit{Large Class} and \textit{Data Clumps}, where it significantly outperforms DeepSeek across languages. Nevertheless, the models are more comparable in certain cases, such as \textit{Switch Statements} in JavaScript and \textit{Shotgun Surgery} in Python, where neither model achieves a dominant advantage.

Overall, this language-level analysis reinforces the trend observed in prior sections—GPT-4.0 consistently achieves higher precision across most code smells and languages, while DeepSeek occasionally compensates with slightly higher recall in specific cases. However, the excessive false positives produced by DeepSeek ultimately reduce its reliability. The results suggest that GPT-4.0 is a more effective option for precise and trustworthy code smell detection across multiple programming languages. Further investigation into model-specific tuning for different languages could enhance detection robustness and minimize weaknesses in recall.

\section{Cost Analysis}
\label{sec:cost_analysis}

The cost of using OpenAI GPT-4.0 and DeepSeek-V3 for automated code smell detection varies significantly due to their fundamentally different pricing models \cite{tang2024large, evstafev2025token}. OpenAI GPT-4.0 follows a token-based pricing model where the cost scales with the number of tokens processed \cite{achiam2023gpt}, while DeepSeek-V3 adopts a complexity-based model \cite{neha2025survey}, charging per request based on task difficulty.

To estimate the cost implications, we consider a 114-line Python script (\texttt{Pizza.py}), which contains an estimated 570 to 1,140 tokens, assuming an average of 5 to 10 tokens per line. Given OpenAI GPT-4.0's pricing at the time of conducting this study of \textbf{\$0.03 USD} per 1,000 input tokens and \textbf{\$0.06 USD} per 1,000 output tokens, and assuming a response length of 250 tokens, the estimated cost for analyzing \texttt{Pizza.py} with OpenAI GPT-4.0 is:

\[
\text{Cost}_{\text{OpenAI GPT-4.0}} = \left(\frac{1,000}{1,000} \times 0.03\right) + \left(\frac{250}{1,000} \times 0.06\right) = 0.0375 \text{ USD}
\]

In contrast, DeepSeek-V3 does not tokenize input explicitly but instead evaluates the code based on structure and patterns \cite{liu2024deepseek}, employing a fixed pricing model based on task complexity. The pricing tiers for DeepSeek-V3 are categorized into:
\begin{itemize}
    \item \textbf{Low complexity}: \textbf{\$0.01 - \$0.02 USD}
    \item \textbf{Intermediate complexity}: \textbf{\$0.03 - \$0.05 USD}
    \item \textbf{High complexity}: \textbf{\$0.10 - \$0.20 USD}
\end{itemize}

Given that the analyzed script falls within the low complexity range, the estimated cost for DeepSeek-V3 is between \textbf{\$0.01 USD} and \textbf{\$0.02 USD}. Table~\ref{tab:cost_comparison} summarizes the estimated costs for both models.

\begin{table}[h]
  \caption{Estimated Cost Comparison of OpenAI GPT-4.0 and DeepSeek-V3 for Code Smell Detection on \texttt{Pizza.py}}
  \label{tab:cost_comparison}
  \centering
  \renewcommand{\arraystretch}{1.2}
  \begin{tabular}{lcc}
    \toprule
    \textbf{Model} & 
    \begin{tabular}[c]{@{}c@{}}\textbf{Pricing}\\\textbf{Model}\end{tabular} & 
    \begin{tabular}[c]{@{}c@{}}\textbf{Estimated Cost}\\\texttt{(Pizza.py)}\end{tabular} \\
    \midrule
    OpenAI GPT-4.0 & \$0.03/1K input + \$0.06/1K output tokens & $\sim$\$0.0375 \\
    DeepSeek-V3 & Complexity-based (low complexity) & $\sim$\$0.01--\$0.02 \\
    \bottomrule
  \end{tabular}
\end{table}

These estimates highlight the cost trade-offs between the two models. OpenAI GPT-4.0's pricing scales linearly with token usage, making it more expensive for larger-scale analyses, whereas DeepSeek-V3 offers fixed and predictable pricing, making it more suitable for frequent, high-volume analysis. While OpenAI GPT-4.0 provides more granular and precise code smell detection, its cost may be a limiting factor for large-scale adoption. DeepSeek-V3, though less precise, offers a cost-effective alternative for broader assessments. The choice between these models depends on the trade-off between accuracy, scalability, and budget constraints. 

It is important to note that these values are approximations and may vary depending on implementation details, actual code complexity, and model-specific behavior.

\section{Comparison of SonarQube and LLM-Based Code Analysis}
\label{sec:sonarqube_comparison}

SonarQube is a well-established static analysis tool that enforces predefined coding standards and best practices, ensuring consistency and compliance with established development guidelines. However, its reliance on manually configured rule sets and pattern matching limits its ability to detect more abstract design flaws or patterns of poor software architecture\cite{wu2024ismell}. In contrast, Large Language Models (LLMs) such as GPT-4.0 and DeepSeek offer a more flexible, semantic-based approach to code analysis, identifying a broader range of architectural and design issues aligned with software engineering principles like Bloaters, Couplers, Change Preventers, Dispensables, and Object-Oriented Abusers.

One of the key differences between the two approaches lies in their adaptability. SonarQube requires predefined rule sets and project-specific configurations, making it less suitable for evolving codebases and unconventional coding practices. LLMs, however, analyze code dynamically using natural language prompts, allowing them to adapt to different programming paradigms and styles. This flexibility, while advantageous, introduces variability in results, making LLM-based analysis less deterministic compared to SonarQube \cite {alves2024detecting}. Another significant distinction is the deterministic versus probabilistic nature of these tools. SonarQube applies strict rule-based analysis, ensuring consistent and explainable results but potentially missing design flaws that fall outside its predefined rules. In contrast, LLMs provide contextual insights, recognizing nuanced patterns that are not explicitly covered by static analysis tools. However, their probabilistic nature can lead to inconsistencies, where the same input may yield different results across multiple runs, impacting reproducibility and trust in their assessments.

Integration with development workflows is another area of differentiation. SonarQube is designed for continuous quality monitoring and integrates seamlessly into CI/CD pipelines, providing structured feedback on coding violations and security vulnerabilities. LLMs, in contrast, are more effective for on-demand assessments, refactoring recommendations, and exploratory code analysis. While AI-driven tools offer deeper insights into architectural and design patterns, they currently lack the automation and stability required for continuous enforcement in software development workflows.False positives and negatives also behave differently in both approaches. SonarQube has a lower false-positive rate for syntax and rule-based violations but may overlook higher-level architectural problems such as excessive coupling or misplaced abstractions. LLMs, on the other hand, can identify broader design flaws but are prone to generating false positives due to their generative nature, particularly when reasoning about potential issues without execution data. Additionally, SonarQube’s strict rule enforcement sometimes results in unnecessary warnings, while LLMs may misinterpret code intent, leading to misleading recommendations.

Scalability and performance are also critical considerations. SonarQube is optimized for large-scale enterprise applications, offering efficient batch processing of codebases. However, its performance may degrade when analyzing complex dependency structures or legacy systems with outdated rules. LLMs require substantial computational resources, particularly when processing large codebases, and their token limitations may necessitate iterative queries, increasing analysis latency. Another challenge with LLM-based analysis is explainability. SonarQube provides structured, well-documented feedback based on established coding standards, making it easier for developers to trace and resolve issues. In contrast, LLM-generated insights often lack transparency, making it difficult to validate their reasoning. This lack of interpretability reduces trust in AI-driven code assessments, particularly in enterprise environments where accountability is crucial.

Given the strengths and weaknesses of both approaches, a hybrid solution may be the most effective for comprehensive code quality assessment. SonarQube can serve as a baseline static analysis tool, enforcing coding standards and detecting well-defined issues, while LLMs can complement it by identifying higher-level architectural flaws and assisting in code reviews. By integrating SonarQube’s structured rule enforcement with LLMs’ contextual analysis, organizations can achieve a more balanced and effective approach to software quality assurance. Future advancements should focus on refining AI-driven analysis for greater consistency, improving explainability, and integrating execution-based verification to reduce false positives. The synergy between deterministic and probabilistic analysis methods could pave the way for more intelligent and adaptable software quality assessment frameworks.

\section{Discussion and Conclusion}
\label{sec:discussion_conclusion}

This study offers a comprehensive benchmark of two large language models—GPT-4.0 and DeepSeek-V3—for the task of automated code smell detection across multiple programming languages. Our analysis demonstrates that GPT-4.0 consistently outperforms DeepSeek-V3 in terms of precision, identifying code smells with significantly fewer false positives. This makes it a more reliable candidate for use in professional development workflows, where minimizing noise and ensuring actionable insights are critical. However, both models exhibit relatively low recall, indicating that a substantial number of actual code smells remain undetected. This limitation is especially pronounced in complex categories such as Change Preventers and certain types of Object-Oriented Abusers, where context and architectural understanding play a crucial role.

At the level of specific smell categories and types, GPT-4.0 proved particularly effective in identifying structural issues like Large Class, Message Chains, and Inappropriate Intimacy. DeepSeek-V3, on the other hand, occasionally achieved higher recall in edge cases but suffered from excessive false positives, limiting its trustworthiness. Our type-level and language-based analyses further reveal that both models’ performance varies depending on the programming language, reflecting the influence of syntax, idioms, and language-specific design patterns on code smell detection. GPT-4.0 maintained a generally strong performance across Java, Python, JavaScript, and C++, reinforcing its potential for cross-language generalization.

Cost analysis highlights an important trade-off between model accuracy and affordability. While GPT-4.0 incurs higher per-query costs due to its token-based pricing, its output quality justifies the expense in use cases where precision is essential. DeepSeek-V3 offers lower, fixed costs and faster response times, making it a pragmatic choice for broader exploratory analysis, albeit with reduced diagnostic precision. The comparative assessment with SonarQube reveals a complementary relationship: while SonarQube excels in deterministic, rule-based detection, it lacks the semantic flexibility needed to uncover nuanced architectural flaws. LLMs, particularly GPT-4.0, show an emerging capacity to fill this gap through contextual, semantic reasoning.

These findings open several directions for future work. One promising avenue is improving recall through prompt engineering, few-shot learning, or fine-tuning LLMs on code smell–specific corpora. Hybrid models that integrate LLM-based predictions with traditional static analyzers could balance precision and coverage more effectively. Additionally, the integration of runtime or execution-based context could enhance detection of smells that manifest only during dynamic interactions. A human-in-the-loop evaluation strategy may also improve interpretability and trust in LLM outputs by involving software developers in feedback and validation loops. Finally, extending the evaluation to real-world, domain-specific codebases could help assess model generalizability and drive the creation of more robust and adaptive detection systems.

In conclusion, while LLMs are not yet ready to replace deterministic tools in all aspects of code quality assurance, their capacity to detect higher-level design issues marks a significant step forward. GPT-4.0, in particular, emerges as a strong candidate for integration into intelligent developer tools, enabling more context-aware, human-like software analysis. As these models continue to evolve, they hold the potential to become indispensable companions in modern software engineering.

\section{Appendices}

\subsection{Type-Language Level Comparison}

Table~\ref{tab:type_language_comparison} presents a detailed comparison of GPT-4.0 and DeepSeek-V3 in detecting code smells across different types, displaying true positives (TP), false positives (FP), false negatives (FN), precision, recall, and F1-score.

\begin{table*}[h]
  \caption{Comparison of GPT-4.0 and DeepSeek-V3 in Code Smell Detection at Type-Language Level}
  \label{tab:type_language_comparison}
  \centering
  \begin{tabular}{l l c c c c c c}
    \toprule
    Model & Code Smell & TP & FP & FN & Precision & Recall & F1-score \\
    \midrule
    GPT-4.0 & Large Class & 21 & 1 & 1 & 0.95 & 0.95 & 0.95 \\
    DeepSeek & Large Class & 0 & 22 & 22 & 0.00 & 0.00 & 0.00 \\
    GPT-4.0 & Primitive Obsession & 5 & 4 & 22 & 0.56 & 0.19 & 0.28 \\
    DeepSeek & Primitive Obsession & 4 & 9 & 23 & 0.31 & 0.15 & 0.20 \\
    GPT-4.0 & Data Clumps & 8 & 11 & 13 & 0.42 & 0.38 & 0.40 \\
    DeepSeek & Data Clumps & 0 & 22 & 21 & 0.00 & 0.00 & 0.00 \\
    GPT-4.0 & Long Parameter List & 8 & 8 & 3 & 0.50 & 0.73 & 0.59 \\
    DeepSeek & Long Parameter List & 6 & 15 & 5 & 0.29 & 0.55 & 0.37 \\
    GPT-4.0 & Switch Statements & 21 & 1 & 4 & 0.95 & 0.84 & 0.89 \\
    DeepSeek & Switch Statements & 20 & 0 & 5 & 1.00 & 0.80 & 0.89 \\
    GPT-4.0 & Temporary Field & 1 & 0 & 8 & 1.00 & 0.11 & 0.20 \\
    DeepSeek & Temporary Field & 2 & 9 & 7 & 0.18 & 0.22 & 0.20 \\
    GPT-4.0 & Shotgun Surgery & 0 & 0 & 85 & 0.00 & 0.00 & 0.00 \\
    DeepSeek & Shotgun Surgery & 0 & 10 & 85 & 0.00 & 0.00 & 0.00 \\
    GPT-4.0 & Divergent Change & 3 & 0 & 63 & 1.00 & 0.05 & 0.09 \\
    DeepSeek & Divergent Change & 0 & 9 & 66 & 0.00 & 0.00 & 0.00 \\
    GPT-4.0 & Parallel Inheritance Hierarchies & 2 & 0 & 42 & 1.00 & 0.05 & 0.09 \\
    DeepSeek & Parallel Inheritance Hierarchies & 0 & 0 & 44 & 0.00 & 0.00 & 0.00 \\
    GPT-4.0 & Speculative Generality & 1 & 0 & 21 & 1.00 & 0.05 & 0.09 \\
    DeepSeek & Speculative Generality & 2 & 36 & 20 & 0.05 & 0.09 & 0.07 \\
    GPT-4.0 & Dead Code & 31 & 6 & 0 & 0.84 & 1.00 & 0.91 \\
    DeepSeek & Dead Code & 31 & 17 & 0 & 0.65 & 1.00 & 0.78 \\
    GPT-4.0 & Message Chains & 21 & 0 & 0 & 1.00 & 1.00 & 1.00 \\
    DeepSeek & Message Chains & 19 & 1 & 2 & 0.95 & 0.90 & 0.93 \\
    GPT-4.0 & Middle Man & 21 & 0 & 0 & 1.00 & 1.00 & 1.00 \\
    DeepSeek & Middle Man & 20 & 0 & 1 & 1.00 & 0.95 & 0.98 \\
    GPT-4.0 & Inappropriate Intimacy & 16 & 4 & 5 & 0.80 & 0.76 & 0.78 \\
    DeepSeek & Inappropriate Intimacy & 5 & 2 & 16 & 0.71 & 0.24 & 0.36 \\
    GPT-4.0 & Long Method & 20 & 1 & 1 & 0.95 & 0.95 & 0.95 \\
    DeepSeek & Long Method & 18 & 0 & 3 & 1.00 & 0.86 & 0.92 \\
    GPT-4.0 & Duplicate Code & 16 & 8 & 0 & 0.67 & 1.00 & 0.80 \\
    DeepSeek & Duplicate Code & 16 & 8 & 0 & 0.67 & 1.00 & 0.80 \\
    GPT-4.0 & Refused Bequest & 0 & 4 & 7 & 0.00 & 0.00 & 0.00 \\
    DeepSeek & Refused Bequest & 3 & 15 & 4 & 0.17 & 0.43 & 0.24 \\
    GPT-4.0 & Data Class & 2 & 0 & 0 & 1.00 & 1.00 & 1.00 \\
    DeepSeek & Data Class & 0 & 10 & 2 & 0.00 & 0.00 & 0.00 \\
    GPT-4.0 & Feature Envy & 0 & 1 & 5 & 0.00 & 0.00 & 0.00 \\
    DeepSeek & Feature Envy & 0 & 10 & 5 & 0.00 & 0.00 & 0.00 \\
    GPT-4.0 & Unnecessary Comments & 0 & 0 & 1 & 0.00 & 0.00 & 0.00 \\
    DeepSeek & Unnecessary Comments & 0 & 0 & 1 & 0.00 & 0.00 & 0.00 \\
    \bottomrule
  \end{tabular}
\end{table*}

\subsection{Code Smell Detection Performance Comparison}

Table~\ref{tab:type_language_comparison} presents a detailed comparison between GPT-4.0 and DeepSeek-V3 in detecting various types of code smells across different programming languages. The results are broken down by model, language, and specific smell type, and include key performance metrics such as true positives (TP), false positives (FP), false negatives (FN), precision, recall, and F1-score. This breakdown enables a fine-grained analysis of each model’s strengths and weaknesses, providing insights into their behavior across both general and nuanced detection cases.

\begin{table}[h]
    \caption{Comparison of GPT-4.0 and DeepSeek-V3 in Code Smell Detection at Type-Language Level - part 1}
  \label{tab:type_language_comparison}
  \centering
  \begin{tabular}{l l l c c c c c c}
    \toprule
    Model & Language & Code Smell & TP & FP & FN & Precision & Recall & F1-score \\
    \midrule
    GPT-4.0 & Python & Large Class & 4 & 1 & 1 & 0.8 & 0.8 & 0.80 \\
    DeepSeek & Python & Large Class & 0 & 5 & 5 & 0.0 & 0.0 & 0.0 \\
    GPT-4.0 & Python & Primitive Obsession & 3 & 0 & 5 & 1.0 & 0.375 & 0.55 \\
    DeepSeek & Python & Primitive Obsession & 3 & 2 & 5 & 0.6 & 0.375 & 0.46 \\
    GPT-4.0 & Python & Data Clumps & 3 & 1 & 2 & 0.75 & 0.6 & 0.67 \\
    DeepSeek & Python & Data Clumps & 0 & 5 & 5 & 0.0 & 0.0 & 0.0 \\
    GPT-4.0 & Python & Long Parameter List & 4 & 0 & 1 & 1.0 & 0.8 & 0.89 \\
    DeepSeek & Python & Long Parameter List & 4 & 1 & 1 & 0.8 & 0.8 & 0.80 \\
    GPT-4.0 & Python & Switch Statements & 5 & 0 & 1 & 1.0 & 0.83 & 0.91 \\
    DeepSeek & Python & Switch Statements & 4 & 0 & 2 & 1.0 & 0.67 & 0.8 \\
    GPT-4.0 & Python & Temporary Field & 1 & 0 & 4 & 1.0 & 0.2 & 0.33 \\
    DeepSeek & Python & Temporary Field & 2 & 2 & 3 & 0.5 & 0.4 & 0.44 \\
    GPT-4.0 & Python & Shotgun Surgery & 0 & 0 & 25 & 0.0 & 0.0 & 0.0 \\
    DeepSeek & Python & Shotgun Surgery & 0 & 1 & 25 & 0.0 & 0.0 & 0.0 \\
    GPT-4.0 & Python & Divergent Change & 3 & 0 & 12 & 1.0 & 0.2 & 0.33 \\
    DeepSeek & Python & Divergent Change & 0 & 2 & 15 & 0.0 & 0.0 & 0.0 \\
    GPT-4.0 & Python & Parallel Inheritance Hierarchies & 2 & 0 & 8 & 1.0 & 0.2 & 0.33 \\
    DeepSeek & Python & Parallel Inheritance Hierarchies & 0 & 0 & 10 & 0.0 & 0.0 & 0.0 \\
    GPT-4.0 & Python & Speculative Generality & 1 & 0 & 4 & 1.0 & 0.2 & 0.33 \\
    DeepSeek & Python & Speculative Generality & 2 & 14 & 3 & 0.125 & 0.4 & 0.19 \\
    GPT-4.0 & Python & Dead Code & 10 & 0 & 0 & 1.0 & 1.0 & 1.0 \\
    DeepSeek & Python & Dead Code & 10 & 3 & 0 & 0.77 & 1.0 & 0.87 \\
    GPT-4.0 & Python & Message Chains & 5 & 0 & 0 & 1.0 & 1.0 & 1.0 \\
    DeepSeek & Python & Message Chains & 4 & 0 & 1 & 1.0 & 0.8 & 0.89 \\
    GPT-4.0 & Python & Middle Man & 5 & 0 & 0 & 1.0 & 1.0 & 1.0 \\
    DeepSeek & Python & Middle Man & 5 & 0 & 0 & 1.0 & 1.0 & 1.0 \\
    GPT-4.0 & Python & Inappropriate Intimacy & 2 & 2 & 3 & 0.5 & 0.4 & 0.44\\
    DeepSeek & Python & Inappropriate Intimacy & 0 & 1 & 5 & 0.0 & 0.0 & 0.0 \\
    GPT-4.0 & Python & Long Method & 5 & 0 & 0 & 1.0 & 1.0 & 1.0 \\
    DeepSeek & Python & Long Method & 4 & 0 & 1 & 1.0 & 0.8 & 0.89 \\
    GPT-4.0 & Python & Duplicate Code & 4 & 2 & 0 & 0.67 & 1.0 & 0.8 \\
    DeepSeek & Python & Duplicate Code & 4 & 1 & 0 & 0.8 & 1.0 & 0.89 \\
    GPT-4.0 & Python & Refused Bequest & 0 & 1 & 4 & 0.0 & 0.0 & 0.0 \\
    DeepSeek & Python & Refused Bequest & 3 & 1 & 1 & 0.75 & 0.75 & 0.75 \\
    GPT-4.0 & Python & Data Class & 1 & 0 & 0 & 1.0 & 1.0 & 1.0 \\
    DeepSeek & Python & Data Class & 0 & 4 & 1 & 0.0 & 0.0 & 0.0 \\
    GPT-4.0 & Python & Feature Envy & 0 & 1 & 2 & 0.0 & 0.0 & 0.0 \\
    DeepSeek & Python & Feature Envy & 0 & 1 & 2 & 0.0 & 0.0 & 0.0 \\
    GPT-4.0 & Python & Unnecessary Comments & 0 & 0 & 1 & 0.0 & 0.0 & 0.0 \\
    DeepSeek & Python & Unnecessary Comments & 0 & 0 & 1 & 0.0 & 0.0 & 0.0 \\
    GPT-4.0 & Java & Large Class & 6 & 0 & 0 & 1.0 & 1.0 & 1.0 \\
    DeepSeek & Java & Large Class & 0 & 6 & 6 & 0.0 & 0.0 & 0.0 \\
    GPT-4.0 & Java & Primitive Obsession & 1 & 1 & 5 & 0.5 & 0.17 & 0.25 \\
    DeepSeek & Java & Primitive Obsession & 0 & 2 & 6 & 0.0 & 0.0 & 0.0 \\
    GPT-4.0 & Java & Data Clumps & 5 & 0 & 1 & 1.0 & 0.83 & 0.91 \\
    DeepSeek & Java & Data Clumps & 0 & 6 & 6 & 0.0 & 0.0 & 0.0 \\
    GPT-4.0 & Java & Long Parameter List & 1 & 2 & 0 & 0.33 & 1.0 & 0.5 \\
    DeepSeek & Java & Long Parameter List & 0 & 5 & 1 & 0.0 & 0.0 & 0.0 \\
    GPT-4.0 & Java & Switch Statements & 5 & 1 & 2 & 0.83 & 0.71 & 0.77 \\
    DeepSeek & Java & Switch Statements & 6 & 0 & 1 & 1.0 & 0.86 & 0.92 \\
    GPT-4.0 & Java & Temporary Field & 0 & 0 & 1 & 0.0 & 0.0 & 0.0 \\
    DeepSeek & Java & Temporary Field & 0 & 2 & 1 & 0.0 & 0.0 & 0.0 \\
    GPT-4.0 & Java & Shotgun Surgery & 0 & 0 & 30 & 0.0 & 0.0 & 0.0 \\
    DeepSeek & Java & Shotgun Surgery & 0 & 4 & 30 & 0.0 & 0.0 & 0.0 \\
    GPT-4.0 & Java & Divergent Change & 0 & 0 & 18 & 0.0 & 0.0 & 0.0 \\
    DeepSeek & Java & Divergent Change & 0 & 2 & 18 & 0.0 & 0.0 & 0.0 \\
    GPT-4.0 & Java & Parallel Inheritance Hierarchies & 0 & 0 & 12 & 0.0 & 0.0 & 0.0 \\
    DeepSeek & Java & Parallel Inheritance Hierarchies & 0 & 0 & 12 & 0.0 & 0.0 & 0.0 \\
    \bottomrule
  \end{tabular}
\end{table}

\begin{table}[h]
    \caption{Comparison of GPT-4.0 and DeepSeek-V3 in Code Smell Detection at Type-Language Level - part 2}
  \label{tab:type_language_comparison}
  \centering
  \begin{tabular}{l l l c c c c c c}
    \toprule
    Model & Language & Code Smell & TP & FP & FN & Precision & Recall & F1-score \\
    \midrule
    GPT-4.0 & Java & Speculative Generality & 0 & 0 & 6 & 0.0 & 0.0 & 0.0 \\
    DeepSeek & Java & Speculative Generality & 0 & 8 & 6 & 0.0 & 0.0 & 0.0 \\
    GPT-4.0 & Java & Dead Code & 8 & 1 & 0 & 0.89 & 1.0 & 0.94 \\
    DeepSeek & Java & Dead Code & 8 & 2 & 0 & 0.8 & 1.0 & 0.89 \\
    GPT-4.0 & Java & Message Chains & 5 & 0 & 0 & 1.0 & 1.0 & 1.0 \\
    DeepSeek & Java & Message Chains & 5 & 0 & 0 & 1.0 & 1.0 & 1.0 \\
    GPT-4.0 & Java & Middle Man & 5 & 0 & 0 & 1.0 & 1.0 & 1.0 \\
    DeepSeek & Java & Middle Man & 5 & 0 & 0 & 1.0 & 1.0 & 1.0 \\
    GPT-4.0 & Java & Inappropriate Intimacy & 4 & 1 & 1 & 0.8 & 0.8 & 0.80 \\
    DeepSeek & Java & Inappropriate Intimacy & 1 & 0 & 4 & 1.0 & 0.2 & 0.33 \\
    GPT-4.0 & Java & Long Method & 5 & 1 & 0 & 0.83 & 1.0 & 0.91 \\
    DeepSeek & Java & Long Method & 5 & 0 & 0 & 1.0 & 1.0 & 1.0 \\
    GPT-4.0 & Java & Duplicate Code & 5 & 1 & 0 & 0.83 & 1.0 & 0.91 \\
    DeepSeek & Java & Duplicate Code & 5 & 1 & 0 & 0.83 & 1.0 & 0.91 \\
    GPT-4.0 & Java & Refused Bequest & 0 & 1 & 1 & 0.0 & 0.0 & 0.0 \\
    DeepSeek & Java & Refused Bequest & 0 & 5 & 1 & 0.0 & 0.0 & 0.0 \\
    GPT-4.0 & Java & Data Class & 0 & 0 & 0 & 0.0 & 0.0 & 0.0 \\
    DeepSeek & Java & Data Class & 0 & 1 & 0 & 0.0 & 0.0 & 0.0 \\
    GPT-4.0 & Java & Feature Envy & 0 & 0 & 1 & 0.0 & 0.0 & 0.0 \\
    DeepSeek & Java & Feature Envy & 0 & 3 & 1 & 0.0 & 0.0 & 0.0 \\
    GPT-4.0 & Java & Unnecessary Comments & 0 & 0 & 0 & 0.0 & 0.0 & 0.0 \\
    DeepSeek & Java & Unnecessary Comments & 0 & 0 & 0 & 0.0 & 0.0 & 0.0 \\
    GPT-4.0 & JavaScript & Large Class & 5 & 0 & 0 & 1.0 & 1.0 & 1.0 \\
    DeepSeek & JavaScript & Large Class & 0 & 5 & 5 & 0.0 & 0.0 & 0.0 \\
    GPT-4.0 & JavaScript & Primitive Obsession & 1 & 1 & 7 & 0.5 & 0.125 & 0.2 \\
    DeepSeek & JavaScript & Primitive Obsession & 1 & 2 & 7 & 0.33 & 0.125 & 0.18 \\
    GPT-4.0 & JavaScript & Data Clumps & 0 & 5 & 5 & 0.0 & 0.0 & 0.0 \\
    DeepSeek & JavaScript & Data Clumps & 0 & 5 & 5 & 0.0 & 0.0 & 0.0 \\
    GPT-4.0 & JavaScript & Long Parameter List & 1 & 5 & 1 & 0.17 & 0.5 & 0.25 \\
    DeepSeek & JavaScript & Long Parameter List & 0 & 5 & 2 & 0.0 & 0.0 & 0.0 \\
    GPT-4.0 & JavaScript & Switch Statements & 5 & 0 & 1 & 1.0 & 0.83 & 0.91 \\
    DeepSeek & JavaScript & Switch Statements & 5 & 0 & 1 & 1.0 & 0.83 & 0.91 \\
    GPT-4.0 & JavaScript & Temporary Field & 0 & 0 & 1 & 0.0 & 0.0 & 0.0 \\
    DeepSeek & JavaScript & Temporary Field & 0 & 5 & 1 & 0.0 & 0.0 & 0.0 \\
    GPT-4.0 & JavaScript & Shotgun Surgery & 0 & 0 & 25 & 0.0 & 0.0 & 0.0 \\
    DeepSeek & JavaScript & Shotgun Surgery & 0 & 3 & 25 & 0.0 & 0.0 & 0.0 \\
    GPT-4.0 & JavaScript & Divergent Change & 0 & 0 & 15 & 0.0 & 0.0 & 0.0 \\
    DeepSeek & JavaScript & Divergent Change & 0 & 2 & 15 & 0.0 & 0.0 & 0.0 \\
    GPT-4.0 & JavaScript & Parallel Inheritance Hierarchies & 0 & 0 & 10 & 0.0 & 0.0 & 0.0 \\
    DeepSeek & JavaScript & Parallel Inheritance Hierarchies & 0 & 0 & 10 & 0.0 & 0.0 & 0.0 \\
    GPT-4.0 & JavaScript & Speculative Generality & 0 & 0 & 5 & 0.0 & 0.0 & 0.0 \\
    DeepSeek & JavaScript & Speculative Generality & 0 & 7 & 5 & 0.0 & 0.0 & 0.0 \\
    GPT-4.0 & JavaScript & Dead Code & 7 & 4 & 0 & 0.64 & 1.0 & 0.78 \\
    DeepSeek & JavaScript & Dead Code & 7 & 7 & 0 & 0.5 & 1.0 & 0.67 \\
    GPT-4.0 & JavaScript & Message Chains & 5 & 0 & 0 & 1.0 & 1.0 & 1.0 \\
    DeepSeek & JavaScript & Message Chains & 5 & 0 & 0 & 1.0 & 1.0 & 1.0 \\
    GPT-4.0 & JavaScript & Middle Man & 5 & 0 & 0 & 1.0 & 1.0 & 1.0 \\
    DeepSeek & JavaScript & Middle Man & 5 & 0 & 0 & 1.0 & 1.0 & 1.0 \\
    GPT-4.0 & JavaScript & Inappropriate Intimacy & 4 & 1 & 1 & 0.8 & 0.8 & 0.80 \\
    DeepSeek & JavaScript & Inappropriate Intimacy & 1 & 1 & 4 & 0.5 & 0.2 & 0.29 \\
    GPT-4.0 & JavaScript & Long Method & 5 & 0 & 0 & 1.0 & 1.0 & 1.0 \\
    DeepSeek & JavaScript & Long Method & 4 & 0 & 1 & 1.0 & 0.8 & 0.89 \\
    GPT-4.0 & JavaScript & Duplicate Code & 4 & 2 & 0 & 0.67 & 1.0 & 0.8 \\
    DeepSeek & JavaScript & Duplicate Code & 4 & 4 & 0 & 0.5 & 1.0 & 0.67 \\
    GPT-4.0 & JavaScript & Refused Bequest & 0 & 1 & 1 & 0.0 & 0.0 & 0.0 \\
    DeepSeek & JavaScript & Refused Bequest & 0 & 3 & 1 & 0.0 & 0.0 & 0.0 \\
    GPT-4.0 & JavaScript & Data Class & 1 & 0 & 0 & 1.0 & 1.0 & 1.0 \\
    DeepSeek & JavaScript & Data Class & 0 & 2 & 1 & 0.0 & 0.0 & 0.0 \\
    \bottomrule
  \end{tabular}
\end{table}

\begin{table}[h]
    \caption{Comparison of GPT-4.0 and DeepSeek-V3 in Code Smell Detection at Type-Language Level - part 3}
  \label{tab:type_language_comparison}
  \centering
  \begin{tabular}{l l l c c c c c c}
    \toprule
    Model & Language & Code Smell & TP & FP & FN & Precision & Recall & F1-score \\
    \midrule
    GPT-4.0 & JavaScript & Feature Envy & 0 & 0 & 1 & 0.0 & 0.0 & 0.0 \\
    DeepSeek & JavaScript & Feature Envy & 0 & 3 & 1 & 0.0 & 0.0 & 0.0 \\
    GPT-4.0 & JavaScript & Unnecessary Comments & 0 & 0 & 0 & 0.0 & 0.0 & 0.0 \\
    DeepSeek & JavaScript & Unnecessary Comments & 0 & 0 & 0 & 0.0 & 0.0 & 0.0 \\
    GPT-4.0 & C++ & Large Class & 6 & 0 & 0 & 1.0 & 1.0 & 1.0 \\
    DeepSeek & C++ & Large Class & 0 & 6 & 6 & 0.0 & 0.0 & 0.0 \\
    GPT-4.0 & C++ & Primitive Obsession & 0 & 2 & 5 & 0.0 & 0.0 & 0.0 \\
    DeepSeek & C++ & Primitive Obsession & 0 & 3 & 5 & 0.0 & 0.0 & 0.0 \\
    GPT-4.0 & C++ & Data Clumps & 0 & 5 & 5 & 0.0 & 0.0 & 0.0 \\
    DeepSeek & C++ & Data Clumps & 0 & 6 & 5 & 0.0 & 0.0 & 0.0 \\
    GPT-4.0 & C++ & Long Parameter List & 2 & 1 & 1 & 0.67 & 0.67 & 0.67 \\
    DeepSeek & C++ & Long Parameter List & 2 & 4 & 1 & 0.33 & 0.67 & 0.44 \\
    GPT-4.0 & C++ & Switch Statements & 6 & 0 & 0 & 1.0 & 1.0 & 1.0 \\
    DeepSeek & C++ & Switch Statements & 5 & 0 & 1 & 1.0 & 0.83 & 0.91 \\
    GPT-4.0 & C++ & Temporary Field & 0 & 0 & 2 & 0.0 & 0.0 & 0.0 \\
    DeepSeek & C++ & Temporary Field & 0 & 0 & 2 & 0.0 & 0.0 & 0.0 \\
    GPT-4.0 & C++ & Shotgun Surgery & 0 & 0 & 5 & 0.0 & 0.0 & 0.0 \\
    DeepSeek & C++ & Shotgun Surgery & 0 & 2 & 5 & 0.0 & 0.0 & 0.0 \\
    GPT-4.0 & C++ & Divergent Change & 0 & 0 & 18 & 0.0 & 0.0 & 0.0 \\
    DeepSeek & C++ & Divergent Change & 0 & 3 & 18 & 0.0 & 0.0 & 0.0 \\
    GPT-4.0 & C++ & Parallel Inheritance Hierarchies & 0 & 0 & 12 & 0.0 & 0.0 & 0.0 \\
    DeepSeek & C++ & Parallel Inheritance Hierarchies & 0 & 0 & 12 & 0.0 & 0.0 & 0.0 \\
    GPT-4.0 & C++ & Speculative Generality & 0 & 0 & 6 & 0.0 & 0.0 & 0.0 \\
    DeepSeek & C++ & Speculative Generality & 0 & 7 & 6 & 0.0 & 0.0 & 0.0 \\
    GPT-4.0 & C++ & Dead Code & 6 & 1 & 0 & 0.86 & 1.0 & 0.92 \\
    DeepSeek & C++ & Dead Code & 6 & 5 & 0 & 0.55 & 1.0 & 0.71 \\
    GPT-4.0 & C++ & Message Chains & 6 & 0 & 0 & 1.0 & 1.0 & 1.0 \\
    DeepSeek & C++ & Message Chains & 5 & 1 & 1 & 0.83 & 0.83 & 0.83 \\
    GPT-4.0 & C++ & Middle Man & 6 & 0 & 0 & 1.0 & 1.0 & 1.0 \\
    DeepSeek & C++ & Middle Man & 5 & 0 & 1 & 1.0 & 0.83 & 0.91 \\
    GPT-4.0 & C++ & Inappropriate Intimacy & 6 & 0 & 0 & 1.0 & 1.0 & 1.0 \\
    DeepSeek & C++ & Inappropriate Intimacy & 3 & 0 & 3 & 1.0 & 0.5 & 0.67 \\
    GPT-4.0 & C++ & Long Method & 5 & 0 & 1 & 1.0 & 0.83 & 0.91 \\
    DeepSeek & C++ & Long Method & 5 & 0 & 1 & 1.0 & 0.83 & 0.91 \\
    GPT-4.0 & C++ & Duplicate Code & 3 & 3 & 0 & 0.5 & 1.0 & 0.67 \\
    DeepSeek & C++ & Duplicate Code & 3 & 2 & 0 & 0.6 & 1.0 & 0.75 \\
    GPT-4.0 & C++ & Refused Bequest & 0 & 1 & 1 & 0.0 & 0.0 & 0.0 \\
    DeepSeek & C++ & Refused Bequest & 0 & 6 & 1 & 0.0 & 0.0 & 0.0 \\
    GPT-4.0 & C++ & Data Class & 0 & 0 & 0 & 0.0 & 0.0 & 0.0 \\
    DeepSeek & C++ & Data Class & 0 & 3 & 0 & 0.0 & 0.0 & 0.0 \\
    GPT-4.0 & C++ & Feature Envy & 0 & 0 & 1 & 0.0 & 0.0 & 0.0 \\
    DeepSeek & C++ & Feature Envy & 0 & 3 & 1 & 0.0 & 0.0 & 0.0 \\
    GPT-4.0 & C++ & Unnecessary Comments & 0 & 0 & 0 & 0.0 & 0.0 & 0.0 \\
    DeepSeek & C++ & Unnecessary Comments & 0 & 0 & 0 & 0.0 & 0.0 & 0.0 \\
    \bottomrule
  \end{tabular}
\end{table}

\bibliographystyle{unsrtnat}
\bibliography{main}  %%% Uncomment

\begin{thebibliography}{27}
\providecommand{\natexlab}[1]{#1}
\providecommand{\url}[1]{\texttt{#1}}
\expandafter\ifx\csname urlstyle\endcsname\relax
  \providecommand{\doi}[1]{doi: #1}\else
  \providecommand{\doi}{doi: \begingroup \urlstyle{rm}\Url}\fi

\bibitem[Sadik et~al.(2023{\natexlab{a}})Sadik, Brulin, and Olhofer]{sadik2023coding}
Ahmed~R Sadik, Sebastian Brulin, and Markus Olhofer.
\newblock Coding by design: Gpt-4 empowers agile model driven development.
\newblock \emph{arXiv preprint arXiv:2310.04304}, 2023{\natexlab{a}}.

\bibitem[Sadik et~al.(2023{\natexlab{b}})Sadik, Ceravola, Joublin, and Patra]{sadik2023analysis}
Ahmed~R Sadik, Antonello Ceravola, Frank Joublin, and Jibesh Patra.
\newblock Analysis of chatgpt on source code.
\newblock \emph{arXiv preprint arXiv:2306.00597}, 2023{\natexlab{b}}.

\bibitem[He et~al.(2024)He, Treude, and Lo]{he2024llm}
Junda He, Christoph Treude, and David Lo.
\newblock Llm-based multi-agent systems for software engineering: Literature review, vision and the road ahead.
\newblock \emph{ACM Transactions on Software Engineering and Methodology}, 2024.

\bibitem[Waseem et~al.(2023)Waseem, Das, Ahmad, Liang, Fehmideh, and Mikkonen]{waseem2023chatgpt}
Muhammad Waseem, Teerath Das, Aakash Ahmad, Peng Liang, Mahdi Fehmideh, and Tommi Mikkonen.
\newblock Chatgpt as a software development bot: a project-based study.
\newblock \emph{arXiv preprint arXiv:2310.13648}, 2023.

\bibitem[Aranda et~al.(2024)Aranda, Oliveira, and Soares]{aranda2024catalog}
Manoel Aranda, Naelson Oliveira, and et~al Soares.
\newblock A catalog of transformations to remove smells from natural language tests.
\newblock In \emph{Proceedings of the 28th International Conference on Evaluation and Assessment in Software Engineering}, pages 7--16, 2024.

\bibitem[Lucas et~al.(2024)Lucas, Gheyi, Soares, Ribeiro, and Machado]{lucas2024evaluating}
Keila Lucas, Rohit Gheyi, Elvys Soares, M{\'a}rcio Ribeiro, and Ivan Machado.
\newblock Evaluating large language models in detecting test smells.
\newblock \emph{arXiv preprint arXiv:2407.19261}, 2024.

\bibitem[Velasco et~al.(2024)Velasco, Rodriguez-Cardenas, Palacio, Alif, and Poshyvanyk]{velasco2024propense}
Alejandro Velasco, Daniel Rodriguez-Cardenas, David~N Palacio, Luftar~Rahman Alif, and Denys Poshyvanyk.
\newblock How propense are large language models at producing code smells? a benchmarking study.
\newblock \emph{arXiv preprint arXiv:2412.18989}, 2024.

\bibitem[Lenarduzzi et~al.(2020)Lenarduzzi, Lomio, Huttunen, and Taibi]{lenarduzzi2020sonarqube}
Valentina Lenarduzzi, Francesco Lomio, Heikki Huttunen, and Davide Taibi.
\newblock Are sonarqube rules inducing bugs?
\newblock In \emph{2020 IEEE 27th international conference on software analysis, evolution and reengineering (SANER)}, pages 501--511. IEEE, 2020.

\bibitem[Hong et~al.(2024)Hong, Lee, Ryu, and Baik]{hong2024code}
Hyunsun Hong, Sungu Lee, Duksan Ryu, and Jongmoon Baik.
\newblock Code smell-guided prompting for llm-based defect prediction in ansible scripts.
\newblock \emph{Journal of Web Engineering}, 23\penalty0 (8):\penalty0 1107--1126, 2024.

\bibitem[Wu et~al.(2024)Wu, Mu, Shi, Guo, Liu, Zhuang, Zhong, and Zhang]{wu2024ismell}
Di~Wu, Fangwen Mu, Lin Shi, Zhaoqiang Guo, Kui Liu, Weiguang Zhuang, Yuqi Zhong, and Li~Zhang.
\newblock ismell: Assembling llms with expert toolsets for code smell detection and refactoring.
\newblock In \emph{Proceedings of the 39th IEEE/ACM International Conference on Automated Software Engineering}, pages 1345--1357, 2024.

\bibitem[Alves et~al.(2024)Alves, Santos, Bezerra, and Machado]{alves2024detecting}
Victor~Anthony Alves, Cristiano Santos, Carla Bezerra, and Ivan Machado.
\newblock Detecting test smells in python test code generated by llm: An empirical study with github copilot.
\newblock In \emph{Simp{\'o}sio Brasileiro de Engenharia de Software (SBES)}, pages 581--587. SBC, 2024.

\bibitem[{Refactoring.Guru}()]{refactoring-guru}
{Refactoring.Guru}.
\newblock Code smells.
\newblock \url{https://refactoring.guru/refactoring/smells}.
\newblock Accessed: 2025-02-07.

\bibitem[Marticorena et~al.(2006)Marticorena, L{\'o}pez, and Crespo]{marticorena2006extending}
Ra{\'u}l Marticorena, Carlos L{\'o}pez, and Yania Crespo.
\newblock Extending a taxonomy of bad code smells with metrics.
\newblock In \emph{Proceedings of 7th International Workshop on Object-Oriented Reengineering (WOOR)}, page~6. Citeseer, 2006.

\bibitem[Gesi et~al.(2022)Gesi, Liu, Li, Ahmed, Nagappan, Lo, de~Almeida, Kochhar, and Bao]{gesi2022code}
Jiri Gesi, Siqi Liu, Jiawei Li, Iftekhar Ahmed, Nachiappan Nagappan, David Lo, Eduardo~Santana de~Almeida, Pavneet~Singh Kochhar, and Lingfeng Bao.
\newblock Code smells in machine learning systems.
\newblock \emph{arXiv preprint arXiv:2203.00803}, 2022.

\bibitem[Tandon et~al.(2024)Tandon, Kumar, and Singh]{tandon2024study}
Stuti Tandon, Vijay Kumar, and VB~Singh.
\newblock Study of code smells: A review and research agenda.
\newblock \emph{International Journal of Mathematical, Engineering \& Management Sciences}, 9\penalty0 (3), 2024.

\bibitem[Walter and Alkhaeir(2016)]{walter2016relationship}
Bartosz Walter and Tarek Alkhaeir.
\newblock The relationship between design patterns and code smells: An exploratory study.
\newblock \emph{Information and Software Technology}, 74:\penalty0 127--142, 2016.

\bibitem[Haque et~al.(2018)Haque, Carver, and Atkison]{haque2018causes}
Md~Shariful Haque, Jeff Carver, and Travis Atkison.
\newblock Causes, impacts, and detection approaches of code smell: a survey.
\newblock In \emph{Proceedings of the 2018 ACM Southeast Conference}, pages 1--8, 2018.

\bibitem[Sadik(2025{\natexlab{a}})]{sadik2025smellycode}
Ahmed~R. Sadik.
\newblock Smelly code dataset - python/java/javascript/c++, 2025{\natexlab{a}}.
\newblock URL \url{https://doi.org/10.5281/zenodo.14989674}.

\bibitem[Sadik(2025{\natexlab{b}})]{sadik2025smellycodedataset}
Ahmed~R. Sadik.
\newblock Smelly code dataset - python/java/javascript/c++, 2025{\natexlab{b}}.
\newblock URL \url{https://github.com/HRI-EU/SmellyCodeDataset}.
\newblock Accessed: 2025-02-07.

\bibitem[Paiva et~al.(2017)Paiva, Damasceno, Figueiredo, and Sant’Anna]{paiva2017evaluation}
Thanis Paiva, Amanda Damasceno, Eduardo Figueiredo, and Cl{\'a}udio Sant’Anna.
\newblock On the evaluation of code smells and detection tools.
\newblock \emph{Journal of Software Engineering Research and Development}, 5:\penalty0 1--28, 2017.

\bibitem[Quba et~al.(2021)Quba, Al~Qaisi, Althunibat, and AlZu’bi]{quba2021software}
Gaith~Y Quba, Hadeel Al~Qaisi, Ahmad Althunibat, and Shadi AlZu’bi.
\newblock Software requirements classification using machine learning algorithm’s.
\newblock In \emph{2021 international conference on information technology (ICIT)}, pages 685--690. IEEE, 2021.

\bibitem[Sahin and Tosun(2019)]{sahin2019conceptual}
Sefa~Eren Sahin and Ayse Tosun.
\newblock A conceptual replication on predicting the severity of software vulnerabilities.
\newblock In \emph{Proceedings of the 23rd International Conference on Evaluation and Assessment in Software Engineering}, pages 244--250, 2019.

\bibitem[Tang et~al.(2024)Tang, Xiao, Li, Fang, Zhang, Fong, Lai, Chui, Chan, Wong, et~al.]{tang2024large}
Yiyi Tang, Ziyan Xiao, Xue Li, Qiwen Fang, Qingpeng Zhang, Daniel Yee~Tak Fong, Francisco Tsz~Tsun Lai, Celine Sze~Ling Chui, Esther Wai~Yin Chan, Ian Chi~Kei Wong, et~al.
\newblock Large language model in medical information extraction from titles and abstracts with prompt engineering strategies: A comparative study of gpt-3.5 and gpt-4.
\newblock \emph{medRxiv}, pages 2024--03, 2024.

\bibitem[Evstafev(2025)]{evstafev2025token}
Evgenii Evstafev.
\newblock Token-hungry, yet precise: Deepseek r1 highlights the need for multi-step reasoning over speed in math.
\newblock \emph{arXiv preprint arXiv:2501.18576}, 2025.

\bibitem[Achiam et~al.(2023)Achiam, Adler, Agarwal, Ahmad, Akkaya, Aleman, Almeida, Altenschmidt, Altman, Anadkat, et~al.]{achiam2023gpt}
Josh Achiam, Steven Adler, Sandhini Agarwal, Lama Ahmad, Ilge Akkaya, Florencia~Leoni Aleman, Diogo Almeida, Janko Altenschmidt, Sam Altman, Shyamal Anadkat, et~al.
\newblock Gpt-4 technical report.
\newblock \emph{arXiv preprint arXiv:2303.08774}, 2023.

\bibitem[Neha and Bhati(2025)]{neha2025survey}
Fnu Neha and Deepshikha Bhati.
\newblock A survey of deepseek models.
\newblock \emph{Authorea Preprints}, 2025.

\bibitem[Liu et~al.(2024)Liu, Feng, Xue, Wang, Wu, Lu, Zhao, Deng, Zhang, Ruan, et~al.]{liu2024deepseek}
Aixin Liu, Bei Feng, Bing Xue, Bingxuan Wang, Bochao Wu, Chengda Lu, Chenggang Zhao, Chengqi Deng, Chenyu Zhang, Chong Ruan, et~al.
\newblock Deepseek-v3 technical report.
\newblock \emph{arXiv preprint arXiv:2412.19437}, 2024.

\end{thebibliography}

\end{document}